\documentclass[twocolumn,notrackchanges]{aastex63}

\received{}
\revised{}
\accepted{}
\submitjournal{ApJ}

\usepackage{amsmath,amssymb,mathtools}
\usepackage[textsize=small]{todonotes}
\usepackage{savesym}
\usepackage{subfigure}
\savesymbol{tablenum}
\usepackage{siunitx}
\restoresymbol{SIX}{tablenum}
\usepackage{subfigure}
\usepackage{enumitem}
\usepackage{booktabs}

\newcommand{\ee}{{\rm e}}

\newcommand{\nablab}{{\boldsymbol{\nabla}}}

\newcommand{\PF}{\mathit{PF}}%
\newcommand{\zeroseven}{{\rm RX J0720.4-3125}}%
\newcommand{\eighteen}{{\rm RX J1856.5-3754}}%

\DeclareSIUnit{\gauss}{G}
\DeclareSIUnit{\year}{yr}
\DeclareSIUnit{\erg}{erg}
\DeclareSIUnit{\eV}{eV}
\DeclareSIUnit{\parsec}{pc}
\shorttitle{Emission from Isolated NSs}
\shortauthors{De Grandis et al.}
\graphicspath{{./}{figures/}}

\begin{document}

\title{X-ray Emission from Isolated Neutron Stars revisited: 3D magnetothermal simulations}

\correspondingauthor{Davide De Grandis}
\email{davide.degrandis@phd.unipd.it}

\author{Davide De Grandis}
\affiliation{Department of Physics and Astronomy, University of Padova, via Marzolo 8, 
I-35131 Padova, Italy}
\affiliation{Mullard Space Science Laboratory, University College London, 
Holmbury St. Mary, Surrey,
RH5 6NT, United Kingdom}

\author{Roberto Taverna}
\affiliation{Department of Mathematics and Physics, University of Roma Tre,
via della Vasca Navale 84, I-00146 Roma, Italy}
\affiliation{Department of Physics and Astronomy, University of Padova, via Marzolo 8, 
I-35131 Padova, Italy}

\author{Roberto Turolla}
\affiliation{Department of Physics and Astronomy, University of Padova, via Marzolo 8, 
I-35131 Padova, Italy}
\affiliation{Mullard Space Science Laboratory, University College London, 
Holmbury St. Mary, Surrey,
RH5 6NT, United Kingdom}

\author{Andrea Gnarini}
\affiliation{Department of Mathematics and Physics, University of Roma Tre,
via della Vasca Navale 84, I-00146 Roma, Italy}

\author{Sergei B. Popov}
\affiliation{Sternberg Astronomical Institute, Lomonosov Moscow State University,  Universitetsky prospekt 13, Moscow 119234, Russia
}
\author{Silvia Zane}
\affiliation{Mullard Space Science Laboratory, University College London,
Holmbury St. Mary, Surrey, RH5 6NT, United Kingdom}

\author{Toby S. Wood}
\affiliation{School of Mathematics and Statistics, Newcastle University, 
Newcastle upon Tyne,
NE1 7RU, United Kingdom}

\begin{abstract}

X-ray emission from the surface of isolated neutron stars (NSs) has been now observed in a variety of sources. The ubiquitous presence of pulsations clearly indicates that thermal photons either come from a limited area, possibly heated by some external mechanism, or from the entire (cooling) surface but with an inhomogeneous temperature distribution. In a NS the thermal map is shaped by the magnetic field topology, since heat flows in the crust mostly along the magnetic field lines. Self-consistent surface thermal maps can hence be produced by simulating the coupled magnetic and thermal evolution of the star. We compute the evolution of the neutron star crust in three dimensions for different initial configurations of the magnetic field and use the ensuing thermal surface maps to derive the spectrum and the pulse profile as seen by an observer at infinity, accounting for general-relativistic effects. In particular, we compare cases with a high degree of symmetry with inherently 3D ones, obtained by adding a quadrupole to the initial dipolar field. Axially symmetric fields result in rather small pulsed fractions ($\lesssim 5\%$), while more complex configurations produce higher pulsed fractions, up to $\sim 25\%$. We find that the spectral properties of our axisymmetric model are close to those of the bright isolated NS \eighteen\ at an evolutionary time comparable with the inferred dynamical age of the source.

\end{abstract}


\section{Introduction} \label{sec:intro}

Neutron stars (NSs) are born very hot in the aftermath of the core-collapse supernova event which brings them into existence, and then progressively cool down through neutrino and photon emission. Their thermal evolution is interwoven with that of their 
magnetic field because of the strong dependence on the temperature of the electric and magnetic conductivities. A detailed study of the NS coupled magnetic and thermal evolution is challenging, not the least on the numerical ground, and has been the focus of several investigations over the past 20 years \cite[see e.g.][for reviews and a historical overview]{transport,Pons_2019,2013PhDT.........7V}. 
Most recent efforts were devoted to extending previous axially symmetric calculations to three spatial dimensions. This was  done first by considering the evolution of the $B$-field only \cite[][]{2016PNAS} and then addressing the coupled
magnetothermal problem \citep{2020NatAs.tmp..215I,2020ApJ...903...40D,Igoshev_2021}.

A realistic modeling of NS cooling is key in understanding their internal structure and composition, ultimately bearing to the equation of state of ultra dense matter. In this respect, the detection and interpretation of radiation coming from the (cooling) surface of an isolated NS can provide a direct measure of the star radius and offer insights on the topology of the crustal magnetic field. Thermal emission has been indeed observed, mostly in the soft X-ray band, from several different classes of NSs, including radio pulsars (PSRs), soft gamma repeaters (SGRs) and anomalous X-ray pulsars (AXPs; the magnetar candidates), central compact objects in supernova remnants (CCOs) and a handful of radio-quiet X-ray pulsars known as the ``magnificent seven'' \cite[M7 or XDINSs; see e.g.][for reviews]{2009ASSL..357...91B,2009ASSL..357..141T,2017ARA&A..55..261K,De_Luca_2017}. Besides, thermal emission has been detected in the soft X-ray transients, NS sources in binary systems which cool down during the quiescent state when accretion is turned off \cite[see e.g.][]{2017JApA...38...49W,2020arXiv201110490P}.

Thermal X-rays from NSs are pulsed, indicating that emission comes from a limited region, as in the case of PSRs and SGRs/AXPs, or from the entire star surface, as in the M7 and possibly some of the CCOs, but with a inhomogeneous temperature distribution. Actually, the most general occurrence is a mixture of the two, with localized ``hot spots'', believed to be produced by some form of extra heating (backflowing currents or dissipation of magnetic energy in the crust), superimposed to the rest of the cooling surface, the emission of which is often (but not always) undetected.
Pulse profiles are in many cases far from sinusoidal and the pulsed fraction can be very large, pointing to complex shapes of the emitting region and/or to some form of beaming.
Extracting meaningful information from spectral and timing data requires an \emph{a priori} knowledge of the surface temperature map and emission properties. A simple fitting with a single component spectral model, say a blackbody, would, in fact, provide only the radiation radius, i.e. an estimate of the size of the emitting region (which not necessarily coincides with the star radius) and a value of the temperature which mostly reflects that of the hotter region(s). 

Three-dimensional magnetothermal models can indeed provide a consistent, detailed surface temperature map at different stages of the NS evolution which can be used to compute synthetic spectra and lightcurves for different emission  mechanisms. These can be then confronted with X-ray (and optical) observations. Such an approach has been applied in the past to different sources, the M7 \cite[][]{2006MNRAS.366..727Z,2011A&A...534A..74H}, the magnetar candidates \cite[e.g.][and references therein]{2020MNRAS.492.5057T} and the PSRs \cite[][]{2019ApJ...872...15R}.

Previous studies, however, dealt with thermal surface distributions which were either derived from a prescribed magnetic topology (usually with high degrees of symmetry, e.g. \citealt{2017MNRAS.464.4390P}), or which was given \emph{a priori} to mimic various crustal heating mechanisms \citep[e.g.][]{2014MNRAS.443...31V}, but with no regard to the actual structure of the $B$-field, as it follows from evolution. Very recently \cite{2020NatAs.tmp..215I} presented 3D magnetothermal simulations of magnetars and computed self-consistent lightcurves for a number of sources in quiescence, although the effects of the heat blanketing envelope (see section \ref{sec:magthermevol}) were accounted for in a simplified way.

Among the diverse classes of isolated NSs the M7 are the most promising target for applying 3D magnetothermal simulations to X-ray spectral modeling. These sources, in fact, exhibit purely thermal emission, most likely originating from the entire star surface, are not variable \cite[with the exception of \zeroseven, see again ][]{2009ASSL..357..141T} and, albeit being sometimes dubbed \emph{X-ray dim isolated neutron stars}, are
quite bright in the soft X-rays. As such, they have been repeatedly observed by all major X-ray telescopes, from {\em ROSAT} to {\em Chandra} and especially {\em XMM-Newton}. 

In this work we present models of 3D magnetothermal evolution of isolated neutron stars and  consistently derive their observable properties (phase-dependent spectra, pulse profiles, pulsed fraction). The plan of the paper is as follows. The physics of magnetothermal evolution and the numerical code used for the simulations, together with the tools used to compute the radiation spectrum at infinity, are reviewed in Section \ref{sec:methods}. In Section \ref{sec:results} we present a set of simulations, both in the case of initially axisymmetric and of more complex magnetic configurations. The properties of the former models in connection with those of the brightest of the M7, \eighteen, are discussed in  Section \ref{sec:1856}. Discussion and conclusions follow in Section \ref{sec:conclusions}.

\section{Methods}\label{sec:methods}
\replaced{In this section we briefly describe the tools used in calculating the magnetothermal evolution of isolated neutron stars (INSs) and the radiation spectrum emitted by their cooling surface.}{
The key ingredients for describing the thermal X-ray emission coming from an isolated NS are its surface temperature distribution, which in turn is dictated by the topology of the magnetic field in the crust, and how radiation propagates to the observer, which is greatly influenced by strong gravitational field. In this section, we lay out the tools we employ to compute the thermal and magnetic structure of the crust and its time evolution, and for calculating the thermal flux detected at infinity.
}

\subsection{Magneto-thermal evolution}\label{sec:magthermevol}

{In a magnetized NS, thermal and magnetic evolution are strongly coupled. However, where the star internal magnetic field (mostly) resides is still an open question. In fact, even though it is commonly accepted that the core of a NS is in some sort of superfluid/superconducting state, it is unclear as yet whether the flux expulsion via the Meissner effect is slow enough to allow for some field to be retained in the core over a significant timescale. This reflects in the diverse strategies adopted in previous investigations, which divide into those considering the magneto-thermal evolution of the whole star and the ones focusing on the crust alone. The latter has been the approach of choice when dealing with computations in more than one spatial dimensions \citep[e.g.][see also \citealt{Pons_2019} for a recent review]{2007A&A...470..303P,2013MNRAS.434..123V,2014PhRvL.112q1101G,2014MNRAS.438.1618G}. In this paper we build on the same line, starting from the model presented by \citet{2015PhRvL.114s1101W} to study the crustal magnetic field in three dimensions, but including the thermal evolution of the crust as well. The same approach has been already used in \citet[][in the following DG+20]{2020ApJ...903...40D} and \citet{2020NatAs.tmp..215I}, to which we refer for more details.

The thermal and magnetic evolution of the crust are then described by the induction equation, in the electron MHD regime (eMHD, i.e. in the assumption that only electrons can move), and the heat balance equation
\begin{align}
 \frac{\partial\mathbf{B}}{\partial t} &= -c\,\nablab\times\left[\sigma^{-1}\mathbf{J} +\frac{1}{\ee cn_e}\mathbf{J}\times{\mathbf{B}}+ \boldsymbol{G}\cdot\nablab T-\frac{\nablab\mu}{\ee}\right] \label{eq:Induction}\\
  C_v\frac{\partial T}{\partial t} &= - \nablab\cdot\left( T\boldsymbol{G}\cdot\mathbf{J} - \boldsymbol{k}\cdot\nablab T - \frac{\mu}{\ee}\mathbf{J}\right) + \mathbf{E}\cdot\mathbf{J}+N_\nu \label{eq:heat}
\end{align}
where $n_e$ is the electron density, $\mathbf{J}=c\nablab\times\mathbf{B}/4\pi$ is the current, $\mu$ the chemical potential, $\boldsymbol{G}$ and $\boldsymbol{k}$ are the thermopower and thermal conductivity tensors and $C_v$ is the heat capacity (per unit volume). The term in square brackets in equation (\ref{eq:Induction}) represents the electric field $\mathbf{E}$, as expressed by the generalized Ohm law and $N_\nu$ accounts for energy losses due to neutrino produced by weak processes in the crust. The explicit form we use for $N_\nu$ is detailed in DG+20. However, neutrino emission from the crust is negligible for temperatures below $\approx\SI{e9}{\kelvin}$ \citep{2001PhR...354....1Y}. Since this condition is met for all the models presented in this work, crustal neutrino losses will be not considered in the following. Equations (\ref{eq:Induction}) and (\ref{eq:heat}), written in dimensionless form, were solved in three spatial dimensions by means of the {\sc parody} code (DG+20).

The (numerical) hindrances of a 3D approach force us to introduce a number of simplifying assumptions concerning the microphysical input. These were already discussed in DG+20 but are summarized in the following for the sake of clarity and completeness. We assume that electrons are completely degenerate, so that
\begin{equation}
    \mu= c\hslash (3\pi^2n_e)^{1/3}; \qquad \boldsymbol{G} = -\frac{\pi^2k_B^2T}{\ee\mu}\delta_{ij}
\end{equation}
where $\hslash$ and $k_B$ are the reduced Planck and the Boltzmann constants. Here we only consider the isotropic part of the thermopower tensor, the so-called Seebeck, term which is responsible for the Biermann battery effect. The (scalar) electric and thermal conductivity are taken to be those of electrons,
\begin{equation}
\sigma=\ee^2c^2\,\frac{n_e\tau(\mu)}{\mu}; \qquad (\boldsymbol{k}^{-1})_{ij}=\frac{3\ee^2}{\pi^2k_b^2T} \left(\frac{1}{\sigma}\delta_{ij}+\frac{\epsilon_{ijk}B_k}{\ee c n_e}\right)
\end{equation}
where $\tau$ is the relaxation time and $\epsilon_{ijk}$ is the Levi-Civita symbol. 
A pure eMHD description of the crust implies that other contributions to the transport properties are neglected; in particular, we do not take into account the isotropic terms in the thermal conductivity due to the lattice and superfluid neutrons. While the former contribution is always subdominant, the latter can become comparable if not greater than the electron one for comparatively high temperatures and densities ($T\approx\SI{e8}{\kelvin}$, $\rho\gtrsim\SI{e12}{\gram\per\cubic\centi\meter}$ \citep[see][]{2009PhRvL.102i1101A}.

The crust is in hydrostatic equilibrium and spherically-symmetric, so that the electron density depends only on the radius, $r$. Following \citet{2004ApJ...609..999C}, we choose the scaling $n_e\propto z^4$, where $z$ is the depth into the crust. More specifically, we take  
\begin{align} \label{eq:density}
\mu(r) &= [1 + (1 - r)/0.0463]^{4/3}
\end{align}
with the chemical potential in units of $\mu_0=\SI{2.9e-5}{\erg}$ and $r$ in units of the star radius. This simple expression has been computed by \citet{2014MNRAS.438.1618G} to provide an acceptable fit to a realistic Fe-Ni crustal model with a chosen impurity parameter $Q\simeq3$ based again on \citet{2004ApJ...609..999C}. The density drops from $\rho(r_\text{core})\approx\SI{e14}{\gram\per\cubic\centi\meter}$ to $\rho(r_\text{crust})\approx\SI{e11}{\gram\per\cubic\centi\meter}$ in a $\SI{1}{\kilo\meter}$ thick crust. Although the solid crust can extend down to $\rho\lesssim\SI{e8}{\gram\per\cubic\centi\meter}$, this shallow region is not included in our grid.

 Moreover, we assume that the relaxation time is constant throughout the crust, $\tau=\SI{9.9e-19}{\second}$. This is adequate in the lower crust, but it becomes questionable in the upper layers, where the contribution by phonons becomes important \citep{transport}. The adopted value, however, is such that the conductivity in the upper crust matches the phonon conductivity at a realistic temperature, $T\approx\SI{e8}{\kelvin}$.

Finally, we take the specific heat to be that of the electrons, $C_v=\tau\mu^2T$. This assumption is quite rough, since the (temperature independent) contribution of the ion lattice is dominant or at least of the same order of magnitude under typical NS conditions \citep[see again][and the discussion in DG+20]{transport}. This choice is dictated essentially by computational requirements, since it makes eq. (\ref{eq:heat}) dependent on $T^2$ only (see again DG+20). However, the large conductivity of the crust greatly suppresses the term containing $C_v$ in eq. (\ref{eq:heat}). This implies that the long-term crustal evolution, such as we are considering here, is not too sensitive to the exact form of the specific heat. We are nevertheless aware that this a major limitation and a more realistic expression for $C_v$ will be necessary in investigating impulsive events, like a sudden, localized release of energy in crust (see the discussion in DG+20).}

Since the field is confined within the crust of the NS, the inner boundary conditions are fixed by the requirement that the radial component of the magnetic field and the tangential component of the electric field vanish at the core-crust interface, $r_c$. Since the contribution of the Hall term is negligible at $r_c$, this is tantamount to ask that $B_r(r_c)=0,\ J_t(r_c)=0$ where $J_t$ is the tangential component of the current.
Moreover, we assume that the magnetosphere can only support currents that are negligible with respect to those inside the crust. The outer boundary condition requires then that the field at the top of the crust matches a potential one.

Following again DG+20 (see their Eq. 23), we adopt a simplified time evolution equation for the (isothermal) core temperature, whereas the temperature on the surface of the star is controlled by the properties of the envelope. This is a geometrically thin layer in which a huge temperature gradient is present, and is treated separately. The surface temperature $T_s$ (i.e. that at the top of the envelope) is related to that at the top of the crust, $T_b$, by \added{the relation}
\begin{equation}\label{eq:tsurf}
T_s(T_b, g, \mathbf{B})=T_s^{(0)}(T_b, g)\,\mathcal{X}(T_b, \mathbf{B})
\end{equation}
where $g$ is the gravitational acceleration at the surface. We used the expressions for $T_s^{(0)}$ calculated by \citet{1983ApJ...272..286G} for an iron envelope, with the magnetic correction $\mathcal{X}(T_b,\mathbf{B})$ given by \citet{envelopes}. \added{The latter term describes how the} heat blanketing effect of the envelope is enhanced by the presence of the magnetic field, which thwarts heat transport as field lines get more parallel to the surface.

The solution of equation (\ref{eq:heat}), coupled with equation (\ref{eq:Induction}) and (\ref{eq:tsurf}), provides the thermal map of the surface at any given time. This can be used to simulate the thermal surface emission once a model for radiative processes is specified.
\added{We remark that magnetothermal evolution is computed without taking into account general relativistic effects. Given the limited thickness of the crust, inclusion of GR would result only in a minor change of the magnetic outer boundary condition which has a negligible overall effect \citep[see][DG+20]{2009A&A...496..207P}. Nevertheless, GR crucially affects light propagation from the surface to the observer as discussed in detail in the next section.}

\subsection{Observed spectrum}\label{sec:spectrum}
In order to simulate the emitted radiation from an isolated NS and to reproduce the spectrum as observed at infinity, we exploit the {\sc idl} ray-tracing code originally developed by \citet[see also \citealt{2015MNRAS.454.3254T,2017MNRAS.464.4390P}]{2006MNRAS.366..727Z}. The code starts dividing the star surface into $N_\Theta\times N_\Phi$ angular patches, each one characterized by the co-latitude $\Theta$ and azimuth $\Phi$ of its center in the observer's frame $XYZ$, with the $Z$-axis along the line-of-sight (LOS) and the $X$-axis in the plane of the LOS and the star spin axis $\boldsymbol{\Omega}$. Photons emitted from the surface and propagating along the LOS are then selected, eventually summing together the contributions coming from all the patches which are into view at a given rotational phase $\gamma$.

GR effects are accounted for assuming that the space-time outside the star is described by the vacuum Schwarzschild solution, including both gravitational red-shift and relativistic ray-bending; as a consequence an area larger than an hemisphere is visible at each rotational phase. 
The phase- and energy-dependent radiation flux at the observer is then given by
\begin{align} \label{eq:raybending1}
    F_\nu(\gamma)&=(1-x)\frac{R^2_{\rm NS}}{D^2}\int_0^{2\pi}{\rm d}{\Phi}\int_0^1I_\nu{\rm d}u^2\,,
\end{align}
where $x\equiv 2GM_{\rm NS}/c^2R_{\rm NS}$, with $M_{\rm NS}$ and $R_{\rm NS}$ the star mass and radius and $c$ the speed of light, $D$ is the source distance and $u\equiv\sin\bar{\Theta}$, with
\begin{align} \label{eq:raybending2}
    \bar{\Theta}&=\int_0^{1/2}\frac{{\rm d}v\sin\Theta}{[(1-x)/4-(1-2vx)v^2\sin^2\Theta]^{1/2}}\,.
\end{align}

The specific intensity $I_\nu$ follows once a model for the surface emissivity is specified. \added{Here, $\nu$ is the frequency as measured by a static observer at the star's surface, whereas the frequency at infinity is $\nu_\infty=\nu\sqrt{1-x}$.} It depends in general on the photon propagation direction as well as on the magnetic co-latitude $\theta$ and azimuth $\phi$, which label the emission point in the star frame $pqt$, with the $t$-axis along the magnetic field axis. If the field is axially symmetric, the magnetic axis trivially coincides with that of the dipole component, $\boldsymbol{b}_{\rm dip}$, whereas in more complex topologies a choice must be made. In the non symmetric cases presented in section \ref{sec:results}, we took the field axis to be along the axis of the quadrupole component, $\boldsymbol{b}_{\rm quad}$. The  two angles $\theta$ and $\phi$ can be expressed in terms of those in the observer's frame, $\Theta$ and $\Phi$, and of $\chi$ and $\xi$ ($\psi$), the angles that the spin axis makes with the LOS and $\boldsymbol{b}_{\rm dip}$ ($\boldsymbol{b}_{\rm quad}$), respectively \cite[see][for more details]{2015MNRAS.454.3254T}. It can be easily shown that it is $\psi=\xi\pm\Theta_\mathrm q$, where $\Theta_\mathrm q$ is the angle between $\boldsymbol{b}_{\rm dip}$ and $\boldsymbol{b}_{\rm quad}$ (see again section \ref{sec:tilted}). For each configuration, corresponding to different values of the angles $\chi$ and $\xi$ ($\psi$) on a $21\times21$ evenly spaced grid, the code finally returns the observed spectrum as a function of the rotational phase (chosen from 30 evenly spaced values).

This approach has been used to study the spectral and polarization properties of thermal radiation emitted from NS sources accounting for realistic surface emission models \cite[a magnetic condensate, a magnetized atmosphere,][]{2016MNRAS.459.3585G,2017MNRAS.465..492M,2020MNRAS.492.5057T}. Here, however, we assume
that the (local) spectrum is an isotropic blackbody, i.e.
\begin{align} \label{eq:bbody}
    I_\nu(\theta,\phi)=\frac{2h}{c^2}\frac{\nu^3}{\exp[h\nu/kT(\theta,\phi)]-1}\,,
\end{align}
where $h$ is the Planck's constant and $T(\theta,\phi)$ is the temperature map produced by {\sc parody}. This (simplifying) choice is motivated because our main aim is to assess the effects of considering self-consistent thermal maps on a number of observable quantities (pulse profiles, phase-resolved spectra), rather then producing detailed predictions for the spectral distribution. Moreover, the overall X-ray spectra of the M7 are better interpreted in terms of (one or two) blackbody(s) (with the presence of broad absorption features at $\approx 100$ eV in a few sources), even in the case of bright sources, like \eighteen\ for which a large number of counts is available \cite[][]{2009ASSL..357..141T,2014A&A...563A..50P}. 

\section{Results}\label{sec:results}
We performed numerical 3D simulations to follow the magnetothermal evolution of an isolated neutron star over some Hall times, defined from the scale values as $\tau_H=4\pi n_0\ee R^2/cB_0\approx\SI{e4}{\year}$, exploring, in particular, different initial conditions for the magnetic field. All models refer to a NS with mass $M_\mathrm{NS}=1.2 M_\odot$ and radius $R_\mathrm{NS}=\SI{12}{\kilo\meter}$. In this phase the evolution is dominated by the non-linear Hall term (see equation \ref{eq:Induction}, second term on the rhs), which, albeit being conservative \citep{2007A&A...470..303P}, can transfer energy between different multipolar components of the $B$-field, until a magnetic configuration dominated by odd modes, the so-called \emph{Hall attractor}, is reached before ohmic dissipation takes over \citep{2014PhRvL.112q1101G}. The latter acts more effectively on higher multipoles, hence the field will eventually go back to a dipole-like configuration. This process, however, acts on the much longer Ohm timescale $\tau_O=4\pi\sigma_0 R^2/c^2\approx\SI{e7}{\year}$, which we do not follow here due to the long computational times required by our 3D code.

\begin{figure}
    \centering
    \includegraphics[width=.5\textwidth]{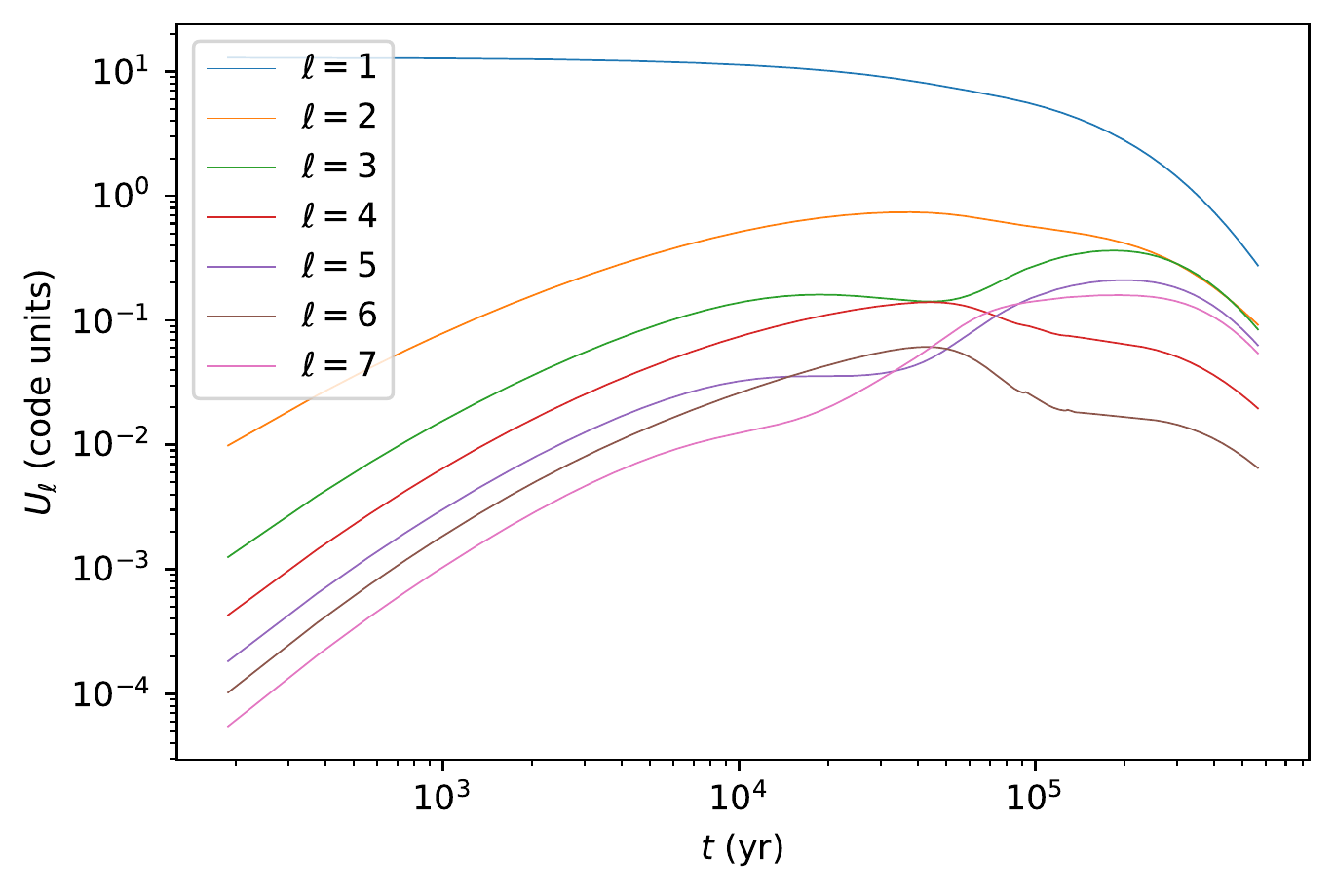}
    \caption{Evolution of the magnetic energy content in the first seven field multipoles for the axially symmetric run discussed in section \ref{sec:axsym}.}
    \label{fig:Hall}
\end{figure} 

The large thermal conductivity of the crust makes equation (\ref{eq:heat}) rather insensitive of the temperature rate of change, so that the star thermal evolution tends to follow the magnetic one: the thermal distribution is influenced by local ohmic dissipation due to currents and by the fact that the electron-mediated heat transport follows the field lines (see {DG+20} and references therein). The magnetic evolution strongly depends on the initial conditions for the $B$-field, so are the latter that actually control the surface thermal map at each evolutionary stage. In this section we present results for a number of cases, starting with axisymmetric initial conditions (section \ref{sec:axsym}) and then moving to more complex magnetic topologies (section \ref{sec:tilted}). 

\subsection{Axisymmetric cases}\label{sec:axsym}
\begin{figure}
    \centering
   \subfigure{\includegraphics[width=.5\textwidth]{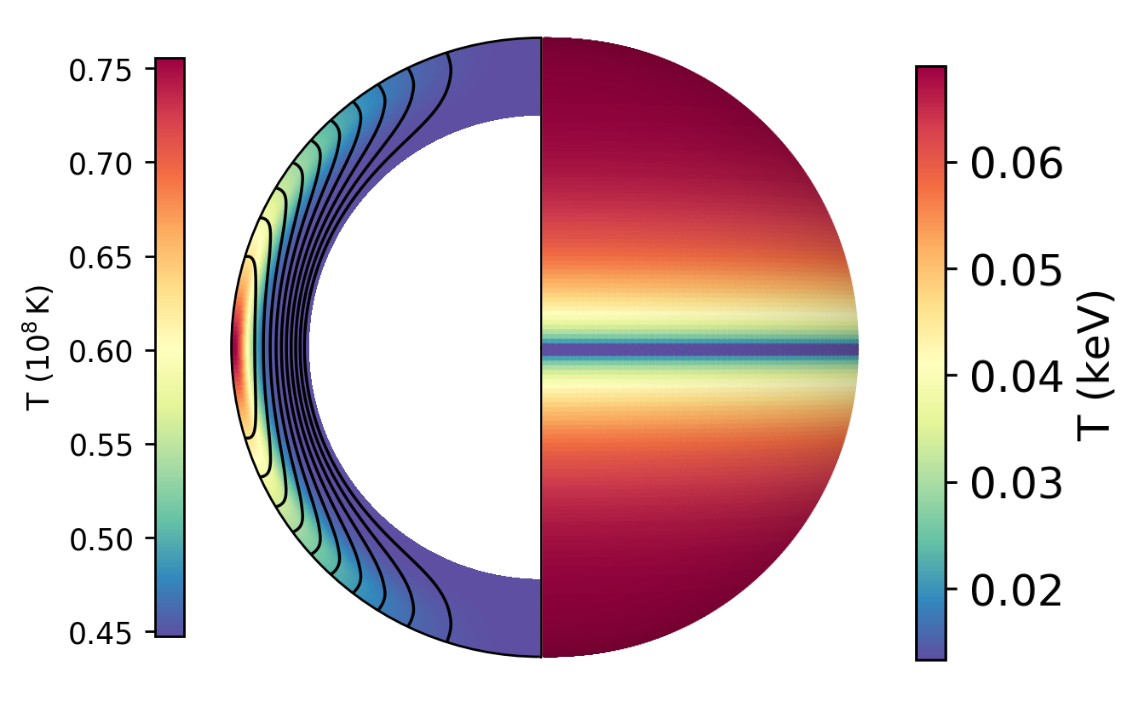}}
   \subfigure{\includegraphics[width=.5\textwidth]{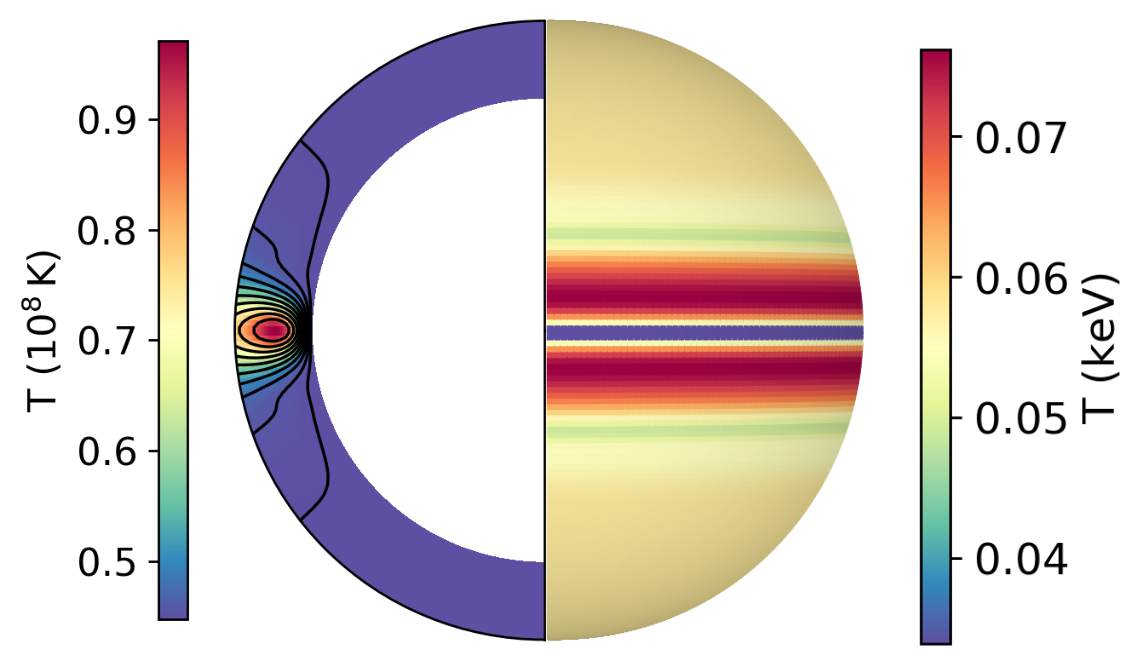}}
    \caption{Internal (left) and surface (right) temperature at an early ($t=\SI{2}{\kilo\year}$, top row) and late ($t=\SI{4e5}{\year}$, bottom row) phase of the evolution. The field lines of the poloidal component are superimposed. The width of the crust is enlarged by a factor $4$ to improve visualization.}
    \label{fig:Tevo}
\end{figure}

We consider first a highly symmetric case, starting the evolution from a force-free, purely dipolar field with $B_\text{dip}(0)\approx\SI{3e13}{\gauss}$, which obeys the boundary conditions. Having previously checked that with such an initial condition the field does not develop instabilities and maintains axial symmetry, as well as north-south (anti)symmetry, during the entire evolution (DG+20), here we restrict the spatial integration domain to a wedge encompassing one sixth of the longitudes with periodic boundary conditions in order to save computational time. Moreover, we decided to reduce the parameter space in this case by suppressing core cooling and setting an initial uniform temperature $T=\SI{4.5e7}{\kelvin}$. This value has been chosen so that to produce a maximum surface temperature of \replaced{$\sim \SI{60}{\eV}$}{$\sim \SI{75}{\eV}$ (corresponding to $\sim \SI{60}{\eV}$ for an observer at infinity)} at an age of $\sim\SI{4e5}{\year}$ in view of the application described in Sec. \ref{sec:1856}. We remark that this, however, does not affect the consistency of our thermal maps, since the core thermal evolution sets the overall temperature scale in the crust, but does not affect its topology, which is dictated by the shape of the magnetic field due again to the large conductivity of the crust.

The evolution of the magnetic field is summarized in Figure \ref{fig:Hall}, where the magnetic energy content of the first seven multipolar components is shown. The formation of the Hall attractor, dominated by odd $\ell$ multipoles, is clearly seen before ohmic dissipation starts to substantially reduce the field at an age of $\approx\SI{e5}{\year}$. 

A general feature of axially symmetric runs is an enhancement of the magnetic field in the equatorial region of the crust, which also becomes hotter. This is due to the currents circulating there and also to the fact that it is threaded by closed field lines, which inhibit heat conduction towards the surface. This hotter internal belt, however, does not automatically translate into a hotter equatorial zone on the surface, since the surface temperature map follows after accounting for heat transport in the envelope. Near the equator, the field lines are almost parallel to the surface, hence the surface temperature is lower, resulting in the appearance of a \emph{colder} equatorial belt.

\begin{figure}
    \centering
    \includegraphics[width=.5\textwidth]{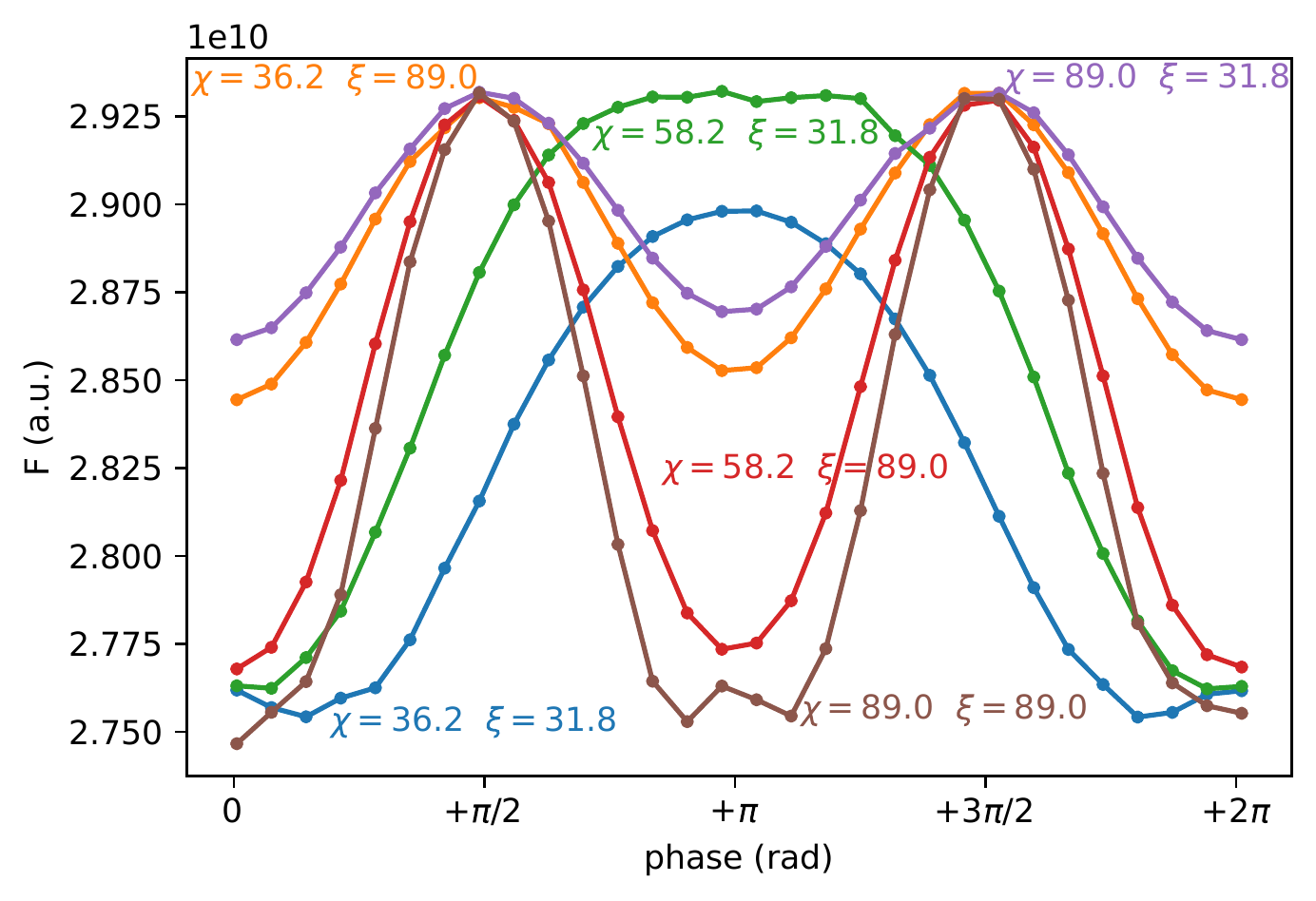}
    \caption{Pulse profiles in the $\num{0.1}-\SI{3}{\kilo\eV}$ 
    energy band for the axisymmetric case at $t=\SI{4e5}{\year}$ for selected values of the two geometrical angles; each curve is labeled by the corresponding values of $\chi$ and $\xi$ (in degrees) and shown in a different color.}
    \label{fig:LCs}
\end{figure}

This is clearly illustrated in Figure \ref{fig:Tevo} where the thermal structure of the crust, together with the temperature map at the top of the crust and at the surface is shown at two subsequent phases of the evolution. In the early phases, the temperature anisotropy is small enough that magnetic screening dominates, and the surface temperature monotonically decreases from the poles to the equator, giving rise to a thermal distribution close to the well-known one produced by a purely dipolar field on the top of an isothermal crust, i.e. a cold equatorial belt and two hotter polar regions \citep[][see also \citealt{1995SSRv...74..437P, 2017MNRAS.464.4390P}]{1983ApJ...271..283G}. On the other hand, as the evolution proceeds, the poloidal field lines are more tightly wound up near the equator and the ``heat trap'' mechanism described previously becomes effective. As a consequence, an equatorial hot region appears on the surface, with a small low temperature ring exactly on the equator, while the polar regions are at an intermediate temperature. 

\begin{figure}
    \centering
    \includegraphics[width=.5\textwidth]{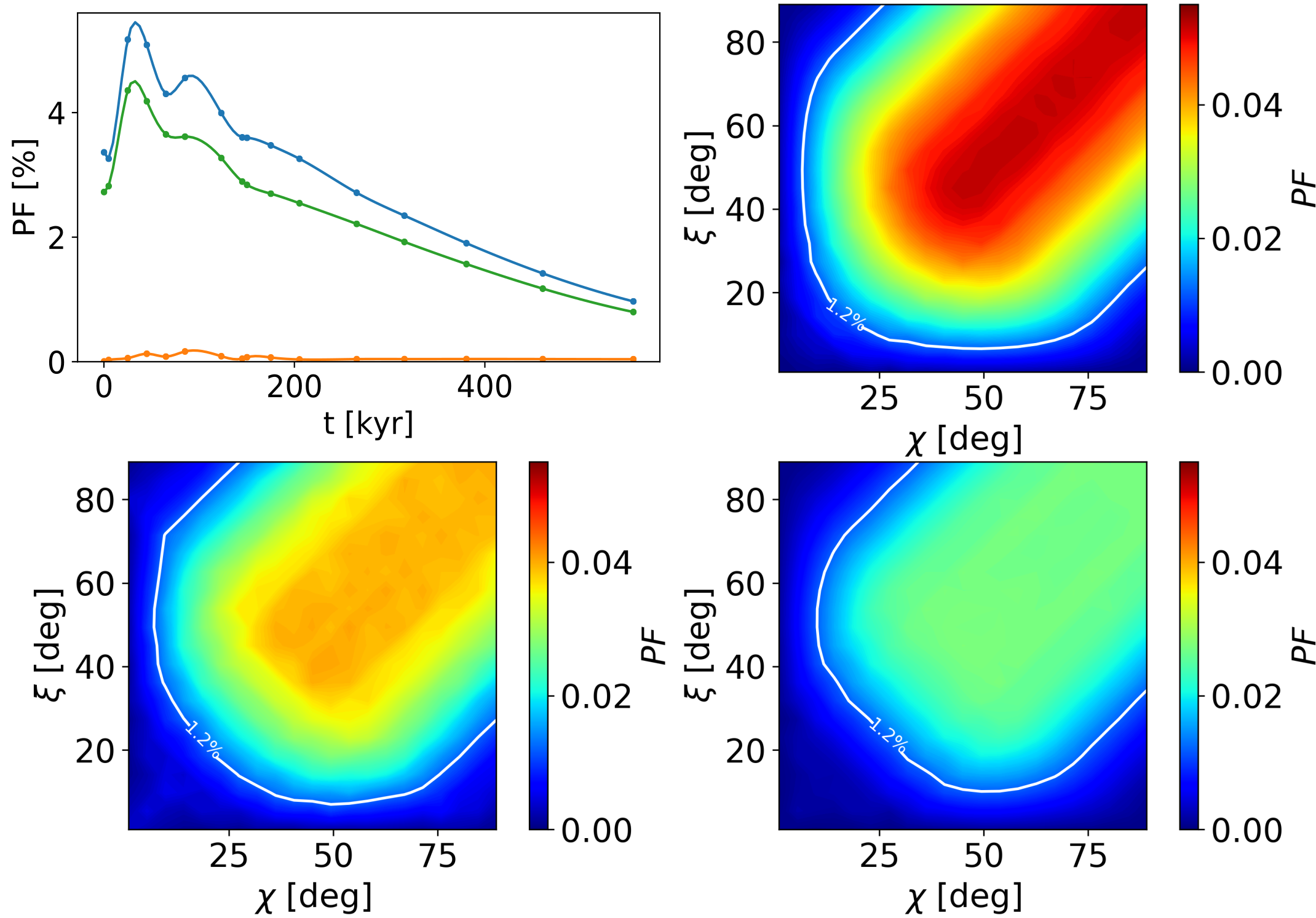}
    \caption{Pulsed fraction as a function of time during the evolution of a NS with axisymmetric field. The three evolution curves correspond to the viewing angles $\chi=\ang{90}$, $\xi=\ang{90}$ (blue, corresponding to the maximum value), $\chi=\ang{36.2}$, $\xi=\ang{31.8}$ (green, see Sec.\ \ref{sec:1856}) and $\chi=\ang{90}$, $\xi=\ang{0}$ (orange). Contour plots show the dependence on the viewing geometry at the selected ages, $\SI{25}{\kilo\year}$ (top-right), $\SI{123}{\kilo\year}$ (bottom-left) and $\SI{265}{\kilo\year}$ (bottom-right). The contour line at $\PF=1.2\%$ corresponds to the value observed for \eighteen\ (see again Sec.\ \ref{sec:1856}).}
    \label{fig:PFevo}
\end{figure}

The thermal maps we obtained were then used as the input for our ray-tracing code in order to get the spectrum received by an observer at infinity as a function of the rotational phase and the two geometrical angles $\chi$ and $\xi$ defined in section \ref{sec:spectrum}. A detailed discussion of the (phase-dependent) spectral distribution is postponed to section \ref{sec:1856}, where an application of our results to the isolated NS \eighteen\ is presented. Here we focus on the pulse profiles and, in particular, on the pulsed fraction and its time evolution. The pulse profiles at an age of $\simeq\SI{4e5}{\year}$ are shown in Figure \ref{fig:LCs} for some representative values of the angles $\chi$ and $\xi$. The pulsation is symmetric around half a period and the profiles are quasi-sinusoidal, with a double-peak structure appearing for those geometries in which the equatorial belt (see Figure \ref{fig:Tevo}) crosses the LOS more than once in a period.

We characterize each lightcurve through its pulsed fraction $\PF$, 
\begin{equation}
    \PF=\frac{F_\text{max}-F_\text{min}}{F_\text{max}+F_\text{min}},
\end{equation}
where $ F_\text{max}$ ($F_\text{min}$) is the maximum (minimum) of the flux, integrated over a given energy band, along the pulse profile. Having specified an emission model (a blackbody in the present case, see \ref{sec:spectrum}), the $\PF$ depends on the thermal map, the viewing geometry and the selected energy range. Fig.\ \ref{fig:PFevo} shows the evolution in time of the maximum $\PF$ (i.e. the one for the most favorable viewing geometry; the minimum one is always $\simeq0$), as well as its dependence on the angles $\chi$ and $\xi$. The $\PF$ peaks at an early stage of the evolution ($t\simeq\SI{25}{\kilo\year}$), when the highest temperature anisotropy in the crust is reached, and then slowly decreases as the crust becomes isothermal and the field is pushed back to a pure dipole by ohmic dissipation. The overall value is anyway quite low, $\PF\lesssim 5\%$. This is expected given the high degree of symmetry of the configuration, and is indeed well below the maximum $\PF$ predicted for an axisymmetric model, which can reach $\sim 10\%$ for comparable star mass and radius for an isotropic emission model \citep{1995SSRv...74..437P}.
The north-south symmetry of the temperature maps is reflected in the invariance of the $\PF$ upon the exchange of the $\chi$ and $\xi$ angles, i.e., the symmetry along the $\chi=\xi$ diagonal in the contour plots of Fig.\ \ref{fig:PFevo}.

\subsubsection{An application: the INS \eighteen} \label{sec:1856}
\eighteen\ was the first source discovered among the M7 and it is the brightest of its class. Since its identification in the Rosat All Sky Survey by \cite{1996Natur.379..233W}, it has been the target of many observations with the {\it XMM-Newton} and {\it Chandra} satellites \citep[see][for a comprehensive analysis of existing data]{2012AA...541A..66S}. Emission \eighteen\ is pulsed at a period of $P\sim 7$ s with a very low pulsed fraction, $\PF\sim 1.2\%$ \citep[][]{2007ApJ...657L.101T}. The lack of an accurate measurement of the period derivative $\dot P$ prevented up to now a robust estimate of the $B$-field: \citet{2008ApJ...673L.163V} provided a spin-down value of $B\sim \SI{1.5e13}{\gauss}$, while a lower value, $B\sim 3-\SI{4e12}{\gauss}$, has been inferred from spectral fitting with atmospheric models by \citet{2007MNRAS.375..821H}. The, quite uncertain, spin-down age of \eighteen \ is  $\tau_\mathrm{c}\approx\SI{4}{\mega\year}$ but dynamical estimates, based on the identification of the NS birthplace, give instead $\tau_\mathrm{dyn}\sim\SI{4e5}{\year}$ \citep[][]{2011MNRAS.417..617T,2013MNRAS.429.3517M}.
\eighteen\ exhibits a purely thermal spectrum, with no absorption features. Analyzing a large set of {\it XMM-Newton} data, \cite{2012AA...541A..66S} found that it is best fitted by two blackbodies, with (redshifted) temperatures $T_1=\SI{62.4}{\eV}$, $T_2=\SI{38.9}{eV}$ ($T_2/T_1\simeq0.59$) and radiation radii $R_1=\SI{4.7}{\kilo\meter}$, $R_2=\SI{11.8}{\kilo\meter}$ at an assumed distance of \SI{120}{\parsec}; this yields a ratio between the emitting areas of $A_2/A_1\simeq6.3$.

\begin{figure}
    \centering
    \includegraphics[width=.5\textwidth]{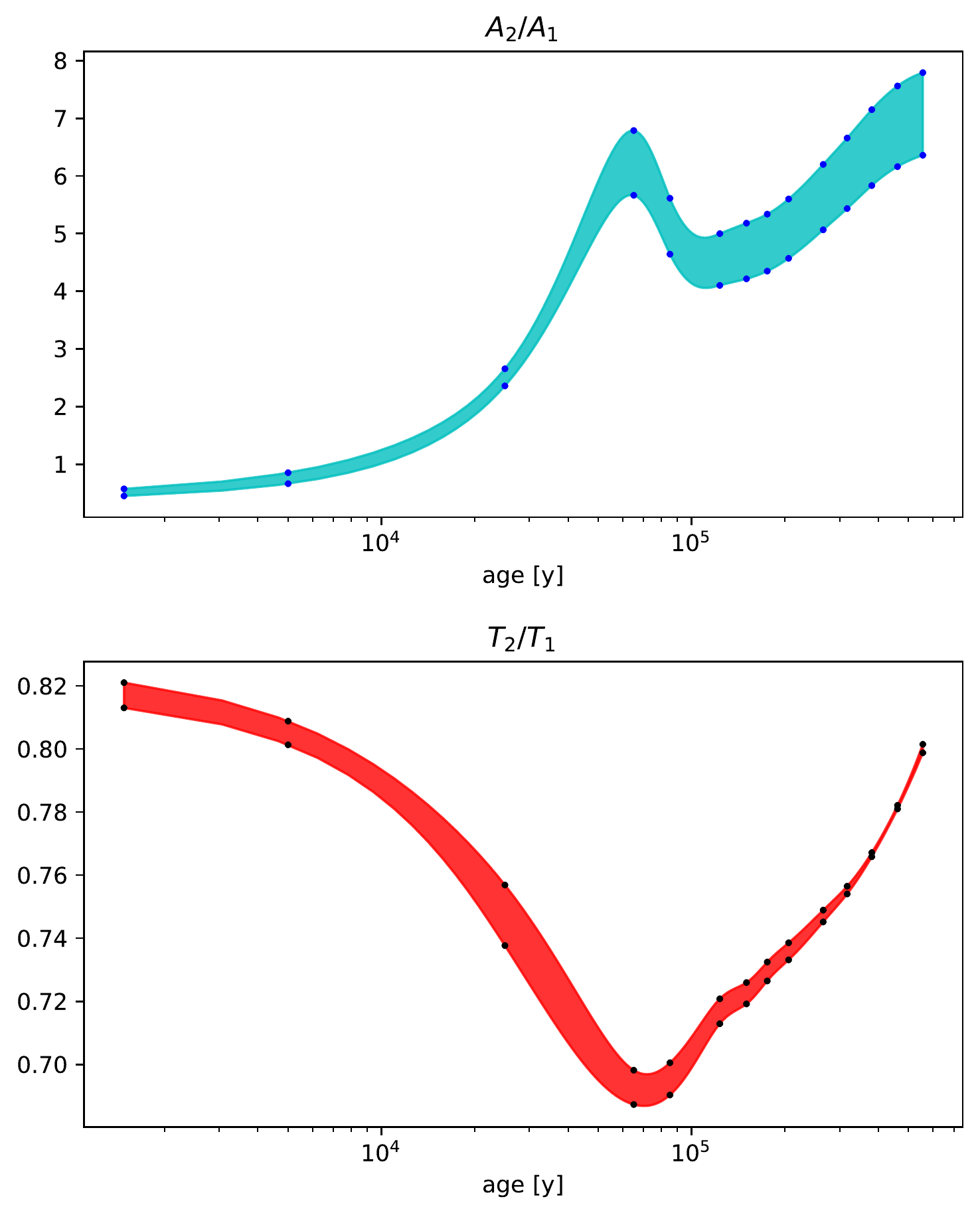}
    \caption{Evolution of the $T_2/T_1$ and $A_2/A_1$ ratios extracted by fitting the synthetic spectra with the double BB  at selected times (results are connected through spline curves). The full ranges of values corresponding to all the viewing geometries explored are reported in the plots.}
    \label{fig:evo_fit}
\end{figure}

This simple spectrum, combined with the very low $\PF\sim1.2\%$, suggest that \eighteen\ harbors a simple, symmetric field. In order to assess if our axisymmetric models can indeed bear to observations, we take run  discussed in section \ref{sec:axsym}, for which the initial conditions were selected to match the dipolar magnetic field (here we take $B_\text{dip}\simeq\SI{4e12}{\gauss}$) and the temperature of the hotter spectral component of \eighteen\ at the inferred dynamical age. We then fitted the computed spectra at different epochs with a double BB function
\begin{equation}\label{eq:2BB}
    F(E)=E^3\left[\displaystyle{\frac{A_1}{\exp(E/T_1)-1}}+
    \displaystyle{\frac{A_2}{\exp(E/T_2) -1}}\right],   
\end{equation}
where $E$ is the photon energy, $T_1$, $T_2$ the temperatures (all in $\si{\kilo\eV}$) and $A_1$, $A_2$ are the emitting areas. 
We find that all our spectra can indeed be interpreted as the superposition of two BB profiles. The ratios $T_2/T_1$ and $A_2/A_1$, as derived from the fits, are shown in Figure \ref{fig:evo_fit} at different epochs. 

\begin{figure}
    \centering
    \includegraphics[width=.5\textwidth]{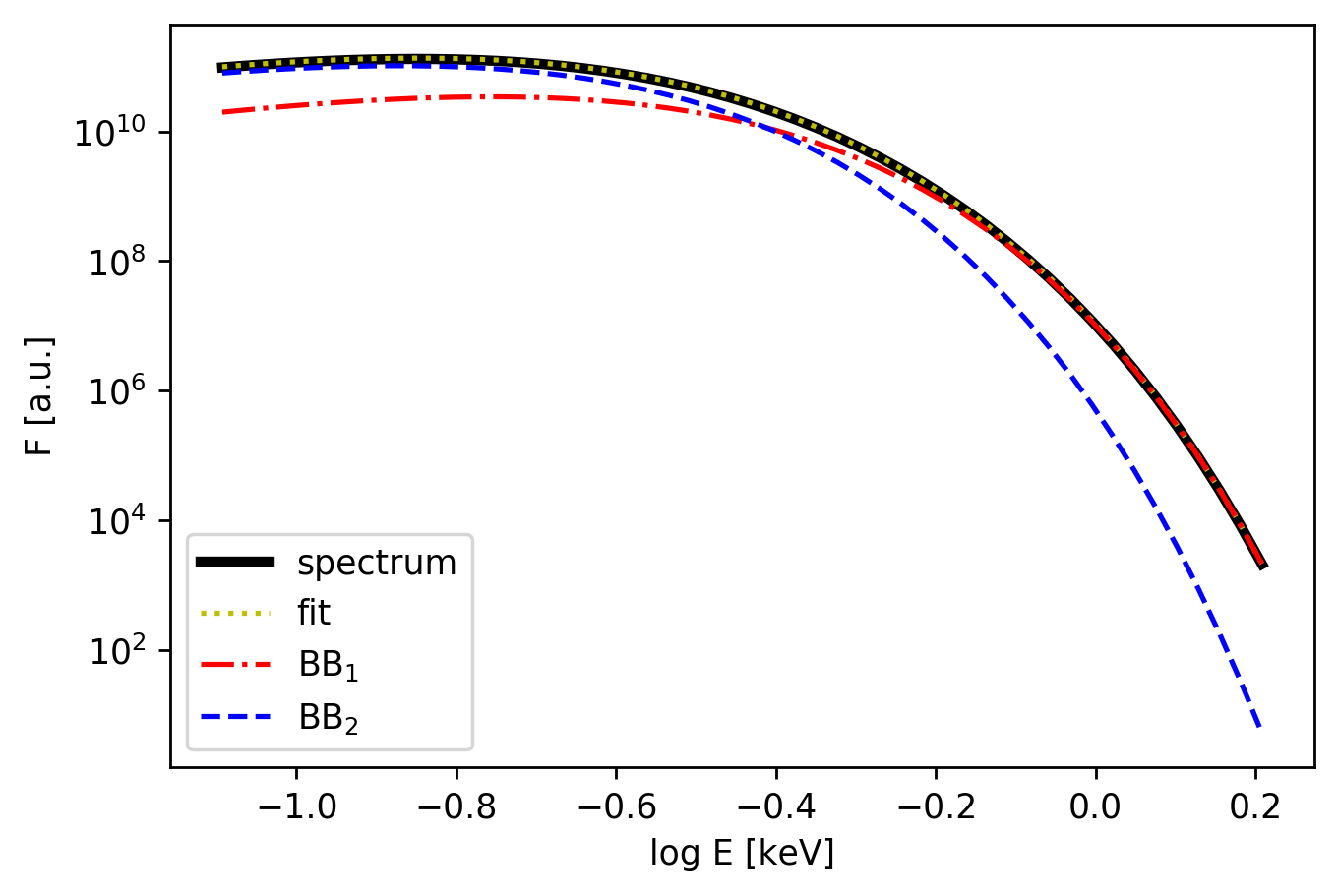}
    \caption{Phase-averaged synthetic spectrum of \eighteen\ (black-solid curve) computed at $t\simeq\SI{4e5}{\year}$ together with the best fitting BB+BB model (yellow-dotted curve, nearly indistinguishable from the calculated spectrum) and the single hotter (red dash-dotted) and hotter (blue dashed) components. The geometrical angles $\chi=\ang{36.2}$ and $\xi=\ang{31.8}$ are chosen in such a way that the observed $\PF\simeq1.2\%$ is reproduced and that the ratio $A_2/A_1=6.3$ is the closest to the observed one.}
    \label{fig:spec1D}
\end{figure}

The fit at a simulation time close to the inferred dynamical age returns a value of  the area of the hotter component which is smaller than that of the colder one. By requiring that the computed pulsed fraction matches the observed one, $\PF\simeq1.2\%$, we get $T_2/T_1\approx0.76$ and $A_2/A_1$ ranging between $6.0$ and $6.7$, according to the selected source geometry. These are indeed in broad agreement with the observations, even if the simplifying assumptions of our model prevent a more quantitative comparison. Figure \ref{fig:spec1D} shows the fit for $\chi=\ang{36.2}$ and $\xi=\ang{31.8}$. This geometry was chosen since it satisfies the $\PF$ requirement and produces a ratio $A_2/A_1=6.3$, which is the closest to observations among the computed ones. 

This result is of interest since earlier attempts to reproduce the spectral parameters of \eighteen\ in the framework of the Hall attractor could reproduce the temperature ratio, but finding consistently that the {\it hot} region had the largest area \citep{2017MNRAS.464.4390P}. This is because only the imprint of the magnetic field on the surface temperature was considered, without taking into account the thermal structure of the crust itself. When considering the effects of the coupling of the magnetic field evolution with temperature in the whole crust, we found that efficient heat trapping mechanism with a high degree of symmetry allows small hot structures to form and persist over long enough timescales.

\subsection{Beyond axial symmetry}\label{sec:tilted}

\begin{table}[]
    \centering
    \begin{tabular}{ccc}
         \toprule
         $\Theta_q$& $\beta$&$\PF_\text{max}$\\
         \midrule
         \ang{45}& 0.75 & 16\%\\
         \ang{45}& 1.25 & 20\%\\
         \ang{45}& 10   & 23\%\\
         \ang{90}& 1    & 4\%\\
         \bottomrule
    \end{tabular}
    \caption{Parameters of the configurations discussed in Sec.\ \ref{sec:tilted}, with the maximum value of the $\PF$ reached in the evolution.}
    \label{tab:rot_cases}
\end{table}

\begin{figure}
    \centering
    \includegraphics[width=.5\textwidth]{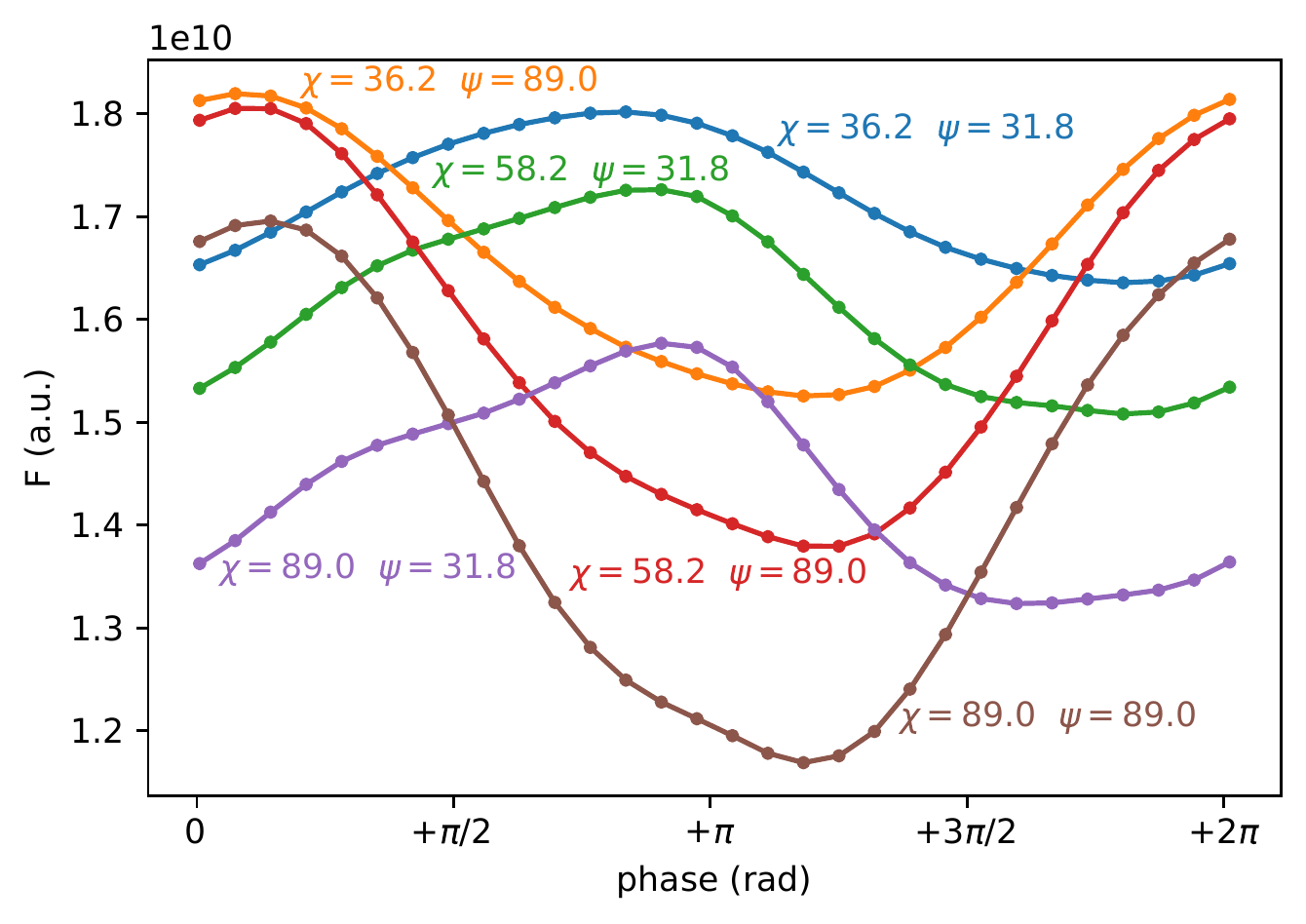}
    \caption{Same as Fig.\ \ref{fig:LCs} for the case $\Theta_q=\ang{45}$, $\beta=1.25$ at time $t=\SI{9570}{\year}$.}
    \label{fig:LCs_rot}
\end{figure}

\begin{figure*}
\centering
\subfigure[Mollweide projection of the thermal maps at the top of the crust (left) and at the surface (right).]{\raisebox{.8em}{\includegraphics[width=.48\textwidth]{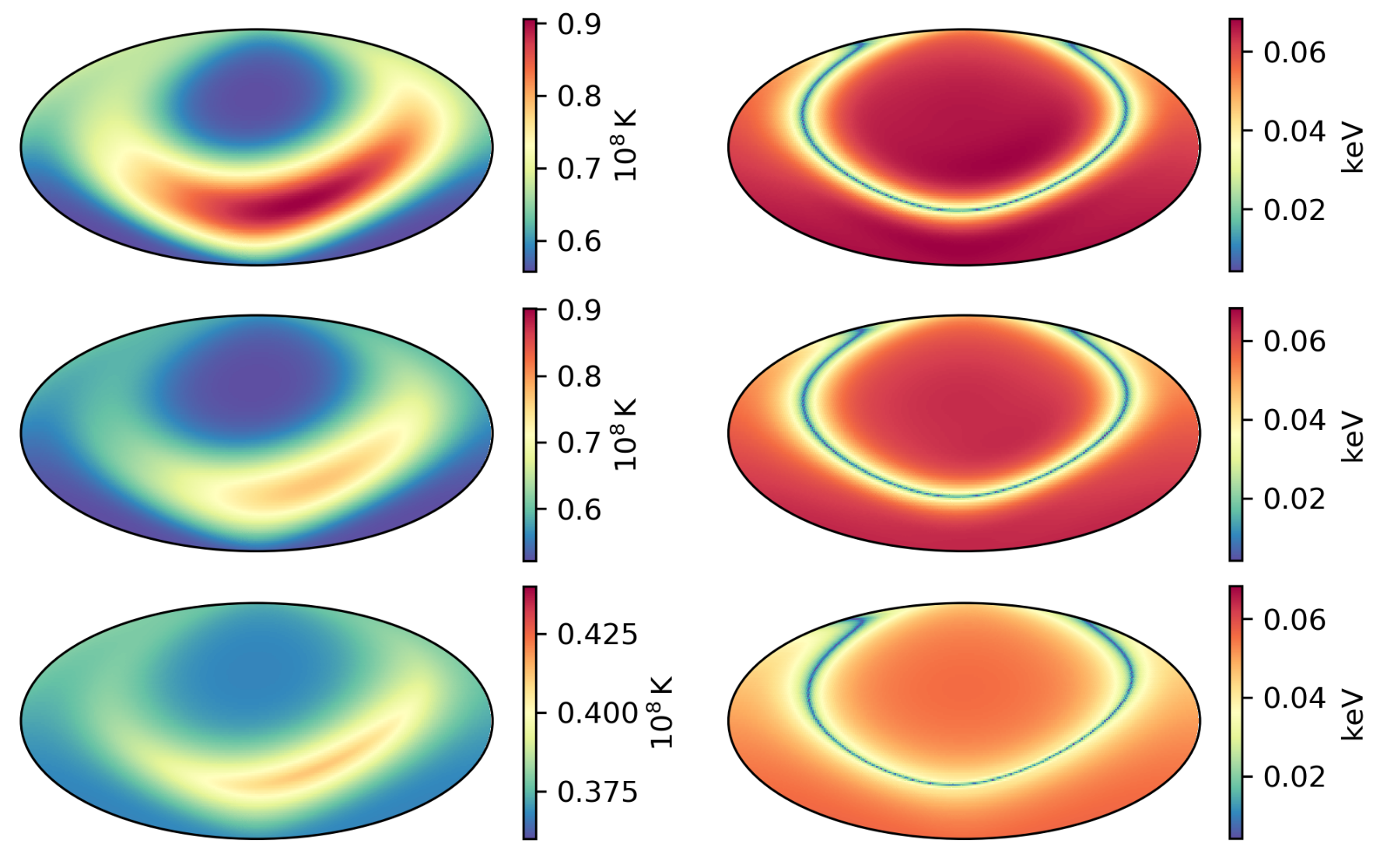}}}~~
\subfigure[Evolution of the $\PF$ for selected viewing angles, alongside its full dependence on $\xi$ and $\psi$ at selected times.] {\includegraphics[width=.49\textwidth]{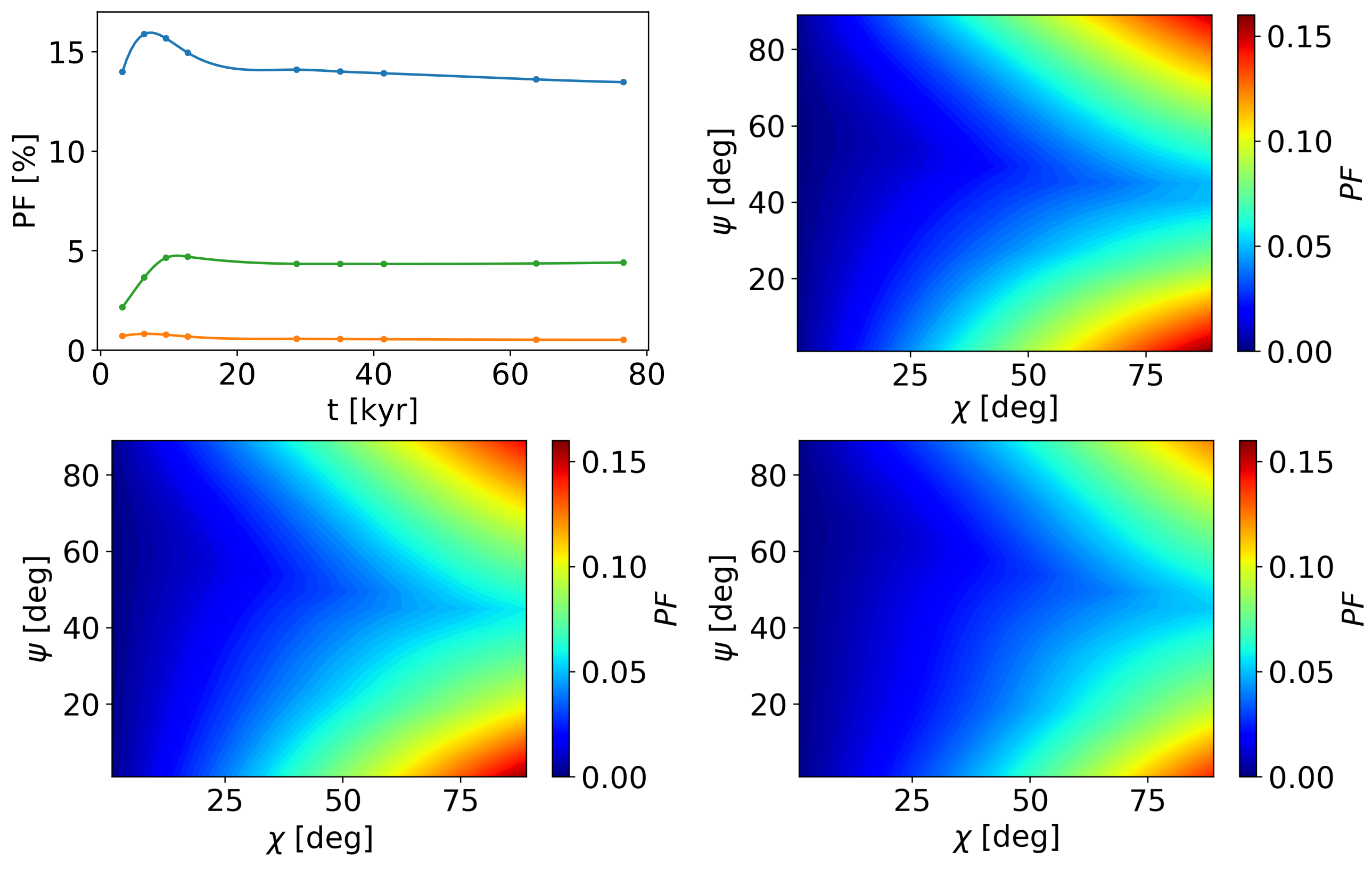}}
\caption{Thermal maps and pulsed fractions for the case $\beta=0.75$, $\Theta_q=\ang{45}$, shown at three times $t_1=\SI{6378}{\year}$, $t_2=\SI{9570}{\year}$ and $t_3=\SI{41476}{\year}$. The $\PF$ evolution curves correspond to the viewing geometries with angles $\chi=\ang{90}$, $\psi=\ang{0}$ (blue, corresponding to the maximum overall value), $\chi=\ang{65}$, $\psi=\ang{45}$ (green) and $\chi=\ang{45}$, $\psi=\ang{45}$ (orange).}\label{fig:rot1}
\end{figure*}

\begin{figure*}
\centering
\subfigure[]{\raisebox{.8em}{\includegraphics[width=.48\textwidth]{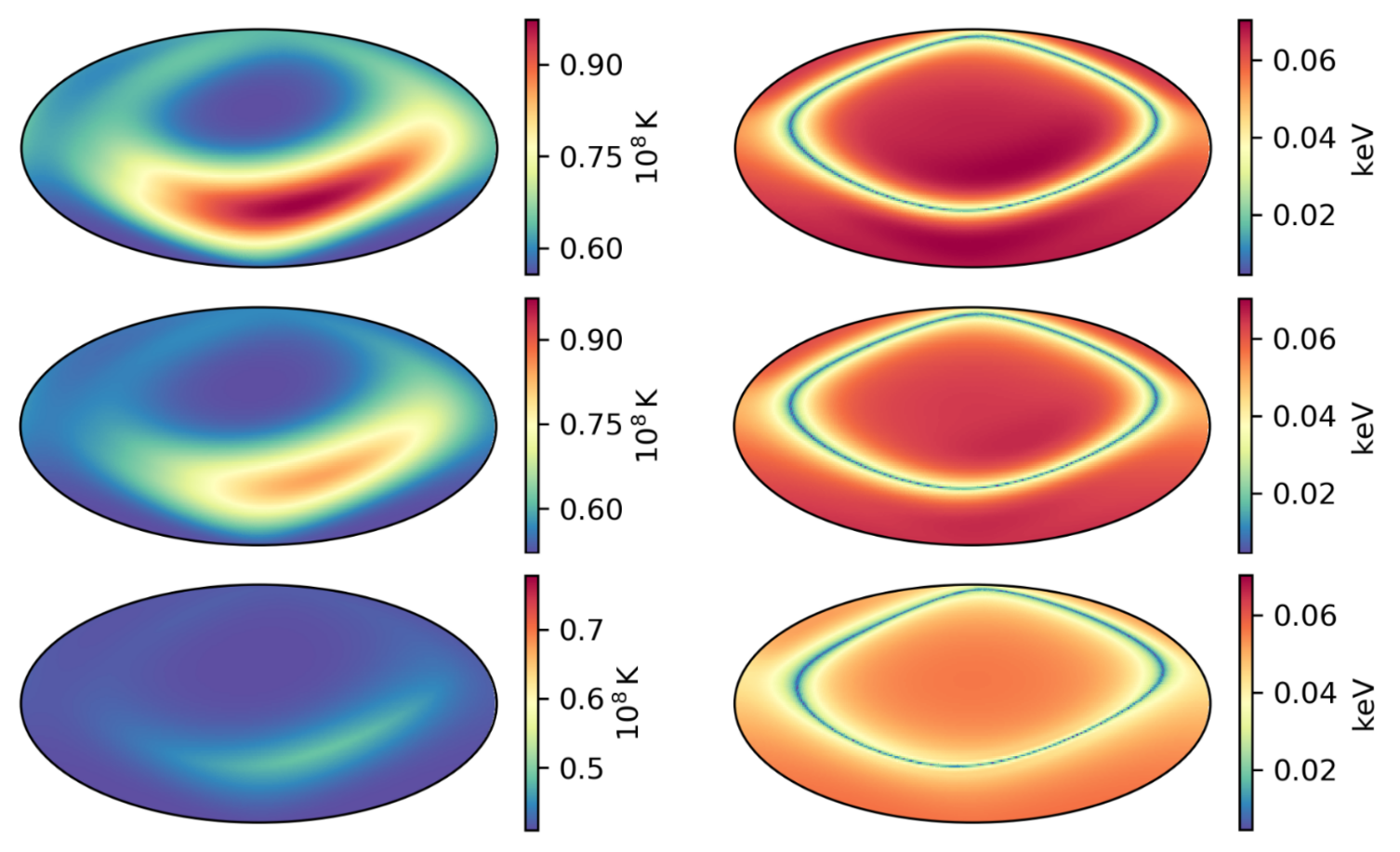}}}~~
\subfigure[]{\includegraphics[width=.49\textwidth]{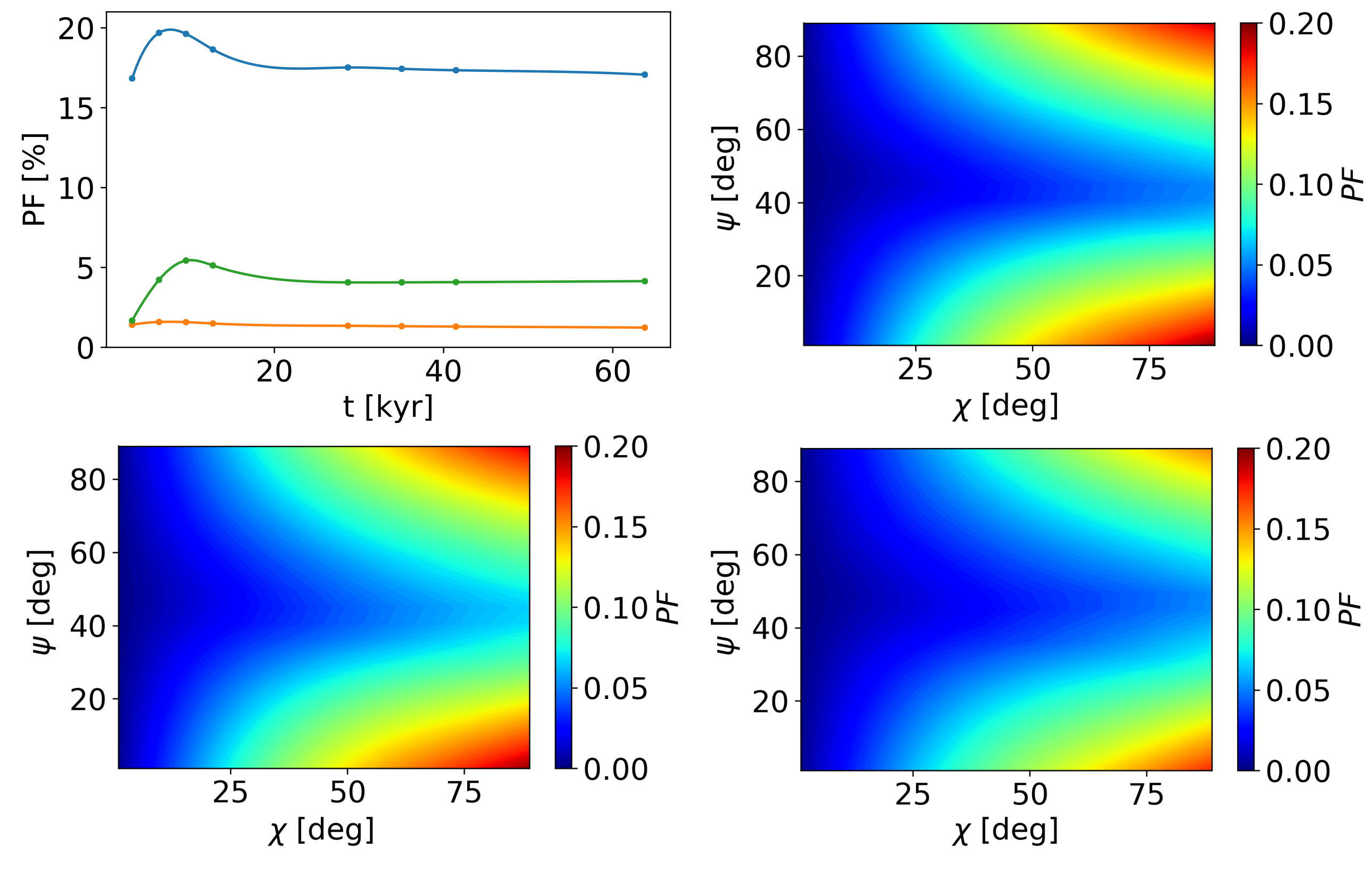}}
\caption{Same as in Figure \ref{fig:rot1} for the case $\beta=1.25$, $\Theta_q=\ang{45}$, shown at three times $t_1=\SI{6378}{\year}$, $t_2=\SI{9570}{\year}$ and $t_3=\SI{41476}{\year}$.}\label{fig:rot2}
\end{figure*}

\begin{figure*}
\centering
\subfigure[]{\raisebox{.8em}{\includegraphics[width=.48\textwidth]{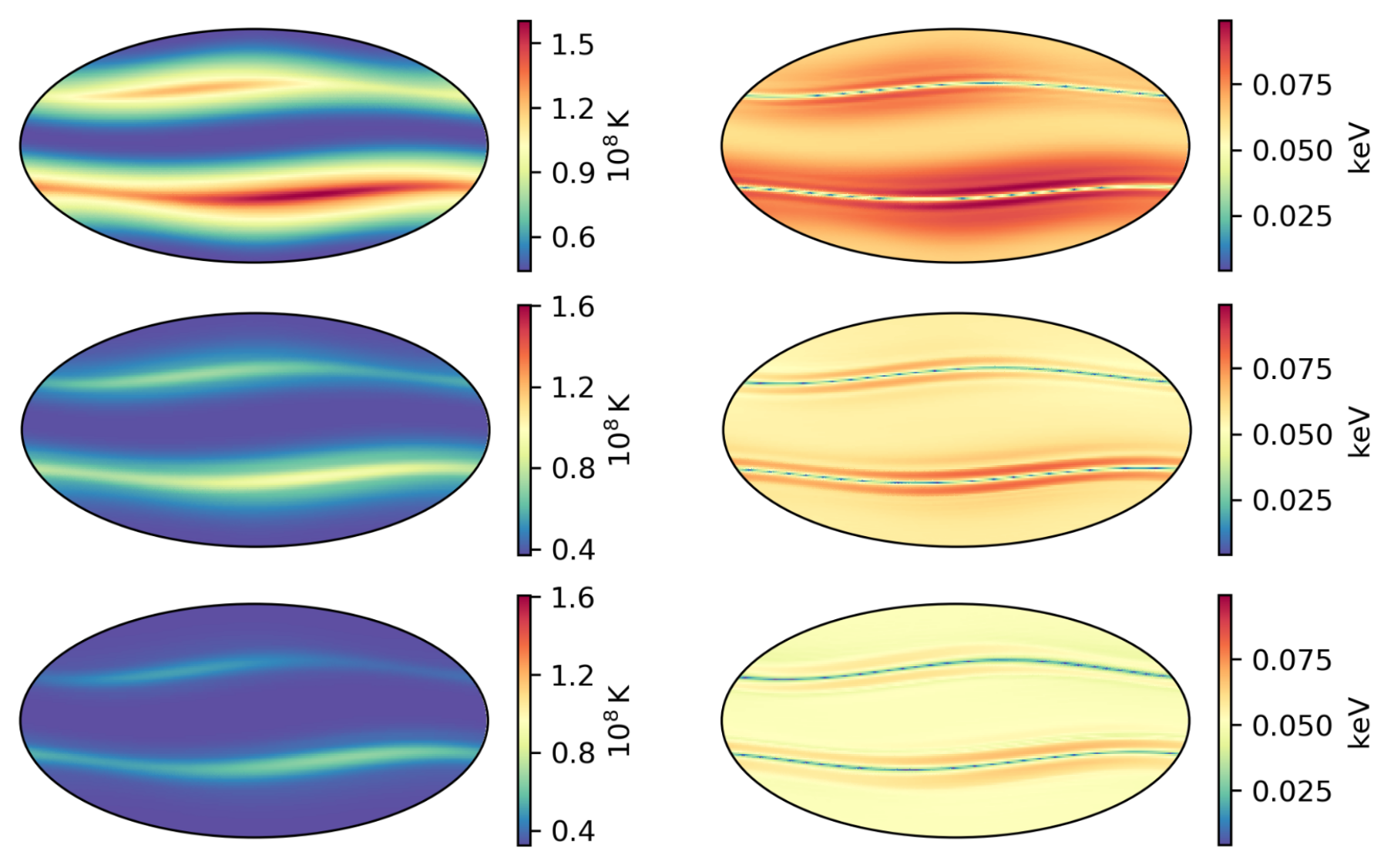}}}~~
\subfigure[]{\includegraphics[width=.49\textwidth]{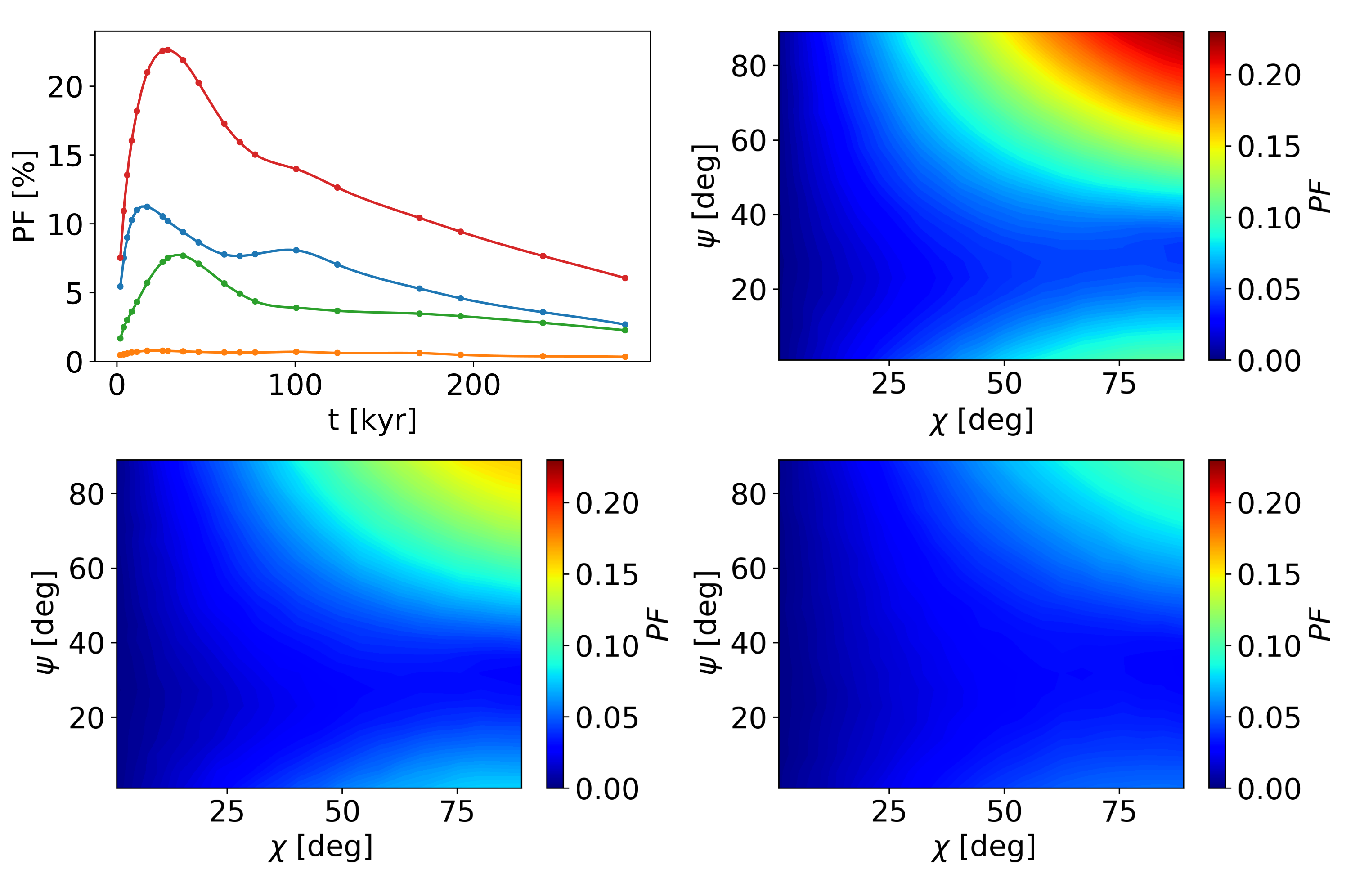}}
\caption{Same as in Figure \ref{fig:rot1} for the case $\beta=10$, $\Theta_q=\ang{45}$, shown at three times $t_1=\SI{25637}{\year}$, $t_2=\SI{68900}{\year}$ and $t_3=\SI{169700}{\year}$ (note that the timescale is quite different from the one of Figs. \ref{fig:rot1} and \ref{fig:rot2}). In this case, the maximum $\PF$ corresponds to the viewing angles $\chi=\ang{90}$, $\psi=\ang{90}$ (red curve).}\label{fig:rot3}
\end{figure*}

\begin{figure*}
\centering
\subfigure[]{\raisebox{.8em}{\includegraphics[width=.48\textwidth]{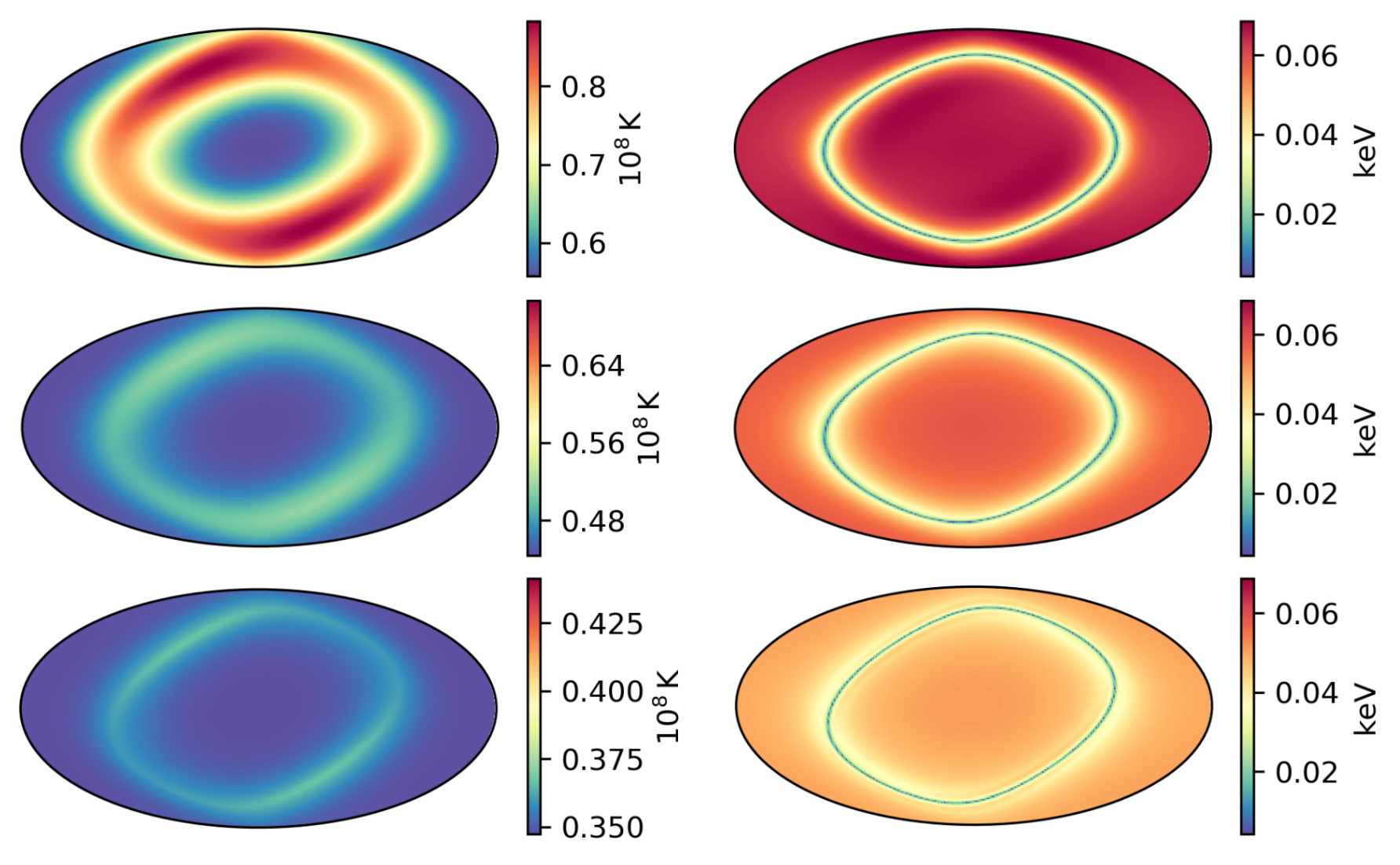}}}~~
\subfigure[]{\includegraphics[width=.49\textwidth]{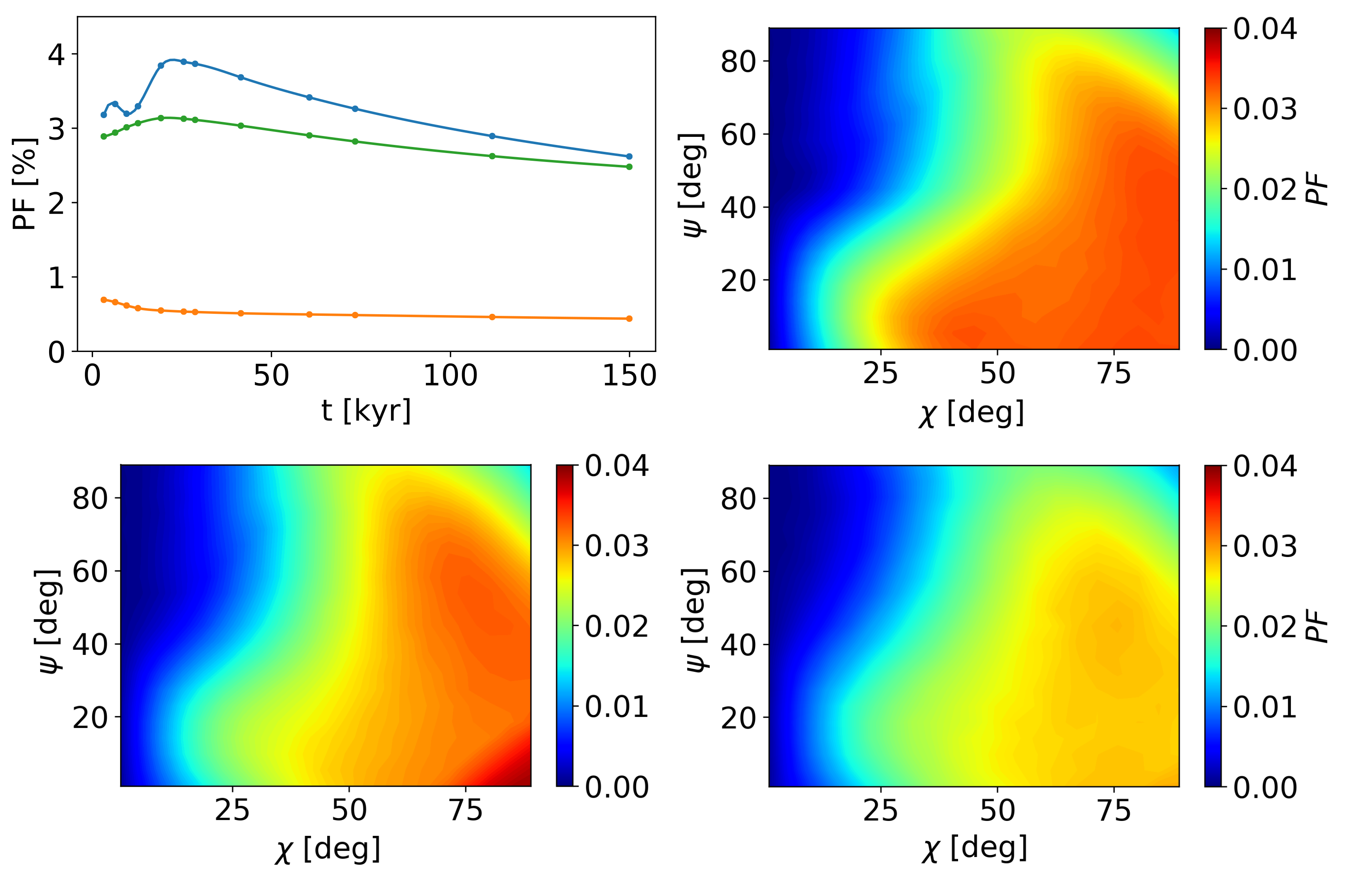}}
\caption{Same as in Figure \ref{fig:rot1} for the case $\beta=1$, $\Theta_q=\ang{90}$, shown at three times $t_1=\SI{6378}{\year}$, $t_2=\SI{68900}{\year}$ and $t_3=\SI{115500}{\year}$ (note that the timescale is quite different from the one of Figs. \ref{fig:rot1} and \ref{fig:rot2}).) }\label{fig:rot4}
\end{figure*}

As we pointed out in the previous section, axially symmetric magnetic configurations produce quasi-sinusoidal pulse profiles with a quite modest pulsed fraction. However, $\PF$ values in excess of $\sim 5\%$ are routinely measured in X-ray pulsars and, in particular, in most of the M7. Indeed, the need for thermal maps produced by more complex field topologies to explain the observed pulse profiles in the M7 was recognized long ago \citep[][]{2006A&A...457..937G,2006MNRAS.366..727Z,2007Ap&SS.308..403P}, although early investigations focused on the structure of the $B$-field in the crust (presence of a toroidal component or higher order multipoles) without addressing its evolution, nor its effects on the temperature of the crust. More recently, \citet{2013MNRAS.434.2362P} presented  surface emission models which were derived self-consistently following the star magnetothermal evolution in 2D (axial symmetry).

In order to address this issue within the framework of 3D magneto-thermal evolution, we now turn to initial field configurations which are not axisymmetric. No attempt to a systematic exploration of the parameter space (which contains a countless number of configurations) will be attempted here. Rather, we consider a limited number of simple cases as illustrative examples and, in particular, we take the initial magnetic field as a combination of (force-free) purely dipolar and quadrupolar poloidal components, with axes tilted by an angle $\Theta_q$. Similar configurations were recently studied by \citet{2020MNRAS.497.2883K} in the context of a detailed stationary-state model of heat transport in the envelope. We tested different angles and relative strengths $\beta=B_\text{quad}(0)/B_\text{dip}(0)$, see Table \ref{tab:rot_cases}, with $B_\text{dip}(0)=\SI{e13}{\gauss}$, $T(0)\equiv\SI{e8}{\kelvin}$. 

The evolution shows that the magnetic field keeps being dominated by these two components, with odd multipoles still favored during the evolution. The angle $\Theta_q$ undergoes virtually no change ($\lesssim\ang{1}$) over the whole time span for all the cases we examined. On the other hand, the thermal structure of the crust is quite different from the axisymmetric cases. In fact, currents still tend to heat the region near the equator (referred to the dipolar component), but in the presence of a non negligible quadrupolar component the field lines are more tangled, so that the heat trapping mechanism is not as efficient. For this reason, the hot region quickly disappears, and the crust becomes almost isothermal over some Hall times.

Considering the first two cases in Table \ref{tab:rot_cases}, the evolution of which is shown in panel (a) of Figures \ref{fig:rot1} and \ref{fig:rot2},  a hotter region  forms on the equator of the dipole (note that the coordinates of the projection are referred to the axis of the quadrupole). However, it is not symmetric, and a spot-like structure appears in one hemisphere only. The resulting surface map is still dominated by magnetic effects, and a colder ring appears in correspondence to the hot region in the crust, whereas a hotter region is present as well, in correspondence to the crustal hot-spot at early times only. 

Now the pulse profiles exhibit a much richer behavior which depends on the degree of asymmetry of the field. An example is shown in Figure \ref{fig:LCs_rot} for the model with $\Theta_q= \ang{45}$, $\beta= 1.25$ at an age of $\sim \SI{10}{\kilo\year}$. As it can be seen, the shape is quite far from sinusoidal for some viewing geometries. 
In Figures \ref{fig:rot1} and \ref{fig:rot2}, panels (b), we show the evolution of the pulsed fraction for various geometries, as well as its dependence on the geometry at selected times. The $\PF$ can reach much higher than in the symmetric cases; at early times ($t\lesssim\SI{10}{\kilo\year}$), is has a peak, after which it decays very slowly. The peak value corresponds to the maximum temperature anisotropy in the crust; then, the evolution of the $\PF$ is dictated by the magnetic field only, on top of an almost isothermal crust. The slow decay can be understood as the effect of ohmic dissipation, which eventually brings the field back to a dipole on longer timescales.

The $\PF$ evolution for these two cases is quite similar, but the one with the larger $\beta$ yields a higher $\PF$. One may, then, expect that a very large quadrupole gives much higher $\PF$s. To test this, we ran a case with $\beta=10$, i.e., close to a pure quadrupole (see Fig.\ \ref{fig:rot3}). Its evolution, however, is very different: the role played by the dipole equator in the thermal evolution is now taken by the two nodes of the quadrupole, and a band structure appears on the surface. Hence, an efficient heat trapping mechanism is in place. This results in longer evolution timescales; in particular, the time required for the crust to become isothermal is longer by almost an order of magnitude with respect to the other configurations. This reflects in an increased amplitude of the $\PF$ peak. Moreover, the range of $\PF$s is wider, and the maximum $\PF$ is just comparable to the one reached with $\beta=1.25$, while values $\lesssim10\%$ are also reached. 

Finally, we tested a different geometry, setting $\Theta_q=\ang{90}$ with $\beta=1$ (see Fig.\ \ref{fig:rot4}). This case is still somewhat symmetric, so that the $\PF$ values are in general low ($\lesssim4\%$), and comparable to the purely symmetric cases. Again, the effect of the quadrupole is to isothermalize the crust, which in this case happens on a timescale $\approx\SI{100}{\kilo\year}$, corresponding to the plateau of the $\PF$ profile. 

\section{Discussion and conclusions}\label{sec:conclusions}

In this work, we presented self-consistent models of the magnetothermal evolution of isolated NSs, focusing in particular on the predicted observable timing and spectral properties at different ages. We have explored the effects of diverse initial magnetic configurations, namely, a purely dipolar field and more complicated topologies, consisting of a dipole plus a (tilted) quadrupole. 

The initial dipolar field evolves, keeping its axial symmetry, towards the so-called \emph{Hall attractor} \citep{2014PhRvL.112q1101G}. As the evolution proceeds, the equatorial region of the crust becomes hotter, owing to the local amplification of the magnetic field and to the formation of closed field lines. The star surface, however, is actually colder near the equator, because heat can not be effectively transported through the envelope where the field lines are nearly parallel to the surface. This leads to the appearance of two hotter bands (one for each magnetic hemisphere) at intermediate latitudes, topped by a colder polar region and joined by an even colder equatorial belt. Assuming that the star surface emits isotropic blackbody radiation at the local temperature, we found that the predicted pulsed fraction is quite small ($\lesssim 5\%$) for all possible orientations of the line-of-sight and of spin axis wrt the magnetic axis. Moreover, the pulsed fraction decreases with age, as the crust becomes more isothermal and ohmic dissipation drives the field back to a dipole. Pulse profiles are both single- and double-peaked, depending on the viewing geometry, and quite sinusoidal (see Section \ref{sec:axsym}).  

The spectrum can be described as a superposition of two blackbody components with different temperatures and emitting areas. This holds at every epoch, although the temperatures and areas vary in time. A quite general feature of these models is that the area of the hotter blackbody component becomes smaller than that of the colder one when the source is older than a few thousands yr. A model with initial dipolar field is in general agreement with the observed properties of the brightest of the M7, \eighteen. The spectral parameters and the pulsed fraction are recovered when the model evolved for  $\sim \SI{4e5}{\year}$, close to the inferred dynamical age of the source, and the dipolar field decayed to about $B_\text{dip}=\SI{4e12}{\gauss}$, a value compatible with that derived by  \citet{2007MNRAS.375..821H} but a factor $2$--$3$ smaller than the spin-down measure. Given the present uncertainties and the hindrance of comparing an inherently multipolar model with estimates derived assuming a dipolar field, we chose to use the value inferred from spectral fitting to investigate the spectral properties. Other runs with different initial field values produce  qualitatively similar results. In particular, the spectrum is always well described by a BB+BB fit with the ratio $A_2/A_1$ eventually getting above unity. Simulations with a much stronger initial field, however, have a high computational cost in the long term, since they not only have to fulfill a more demanding Courant condition, but tend to develop instabilities that call for extremely fine grids.

As we already pointed out, highly symmetrical configurations imply rather low ($\lesssim5\%$) pulsed fractions. Even restricting to NS sources in which emission reasonably comes from the entire star surface, the measured pulsed fraction can be fairly larger: in the M7 it reaches $\sim 30\%$. This suggests that the thermal map in these sources reflects  a more complex (e.g. non-axially symmetric) field configuration. To test this, we considered models in which the initial field consists of a dipole plus a misaligned quadrupole field (the simplest choice for higher order multipole). Indeed, these models exhibit a pulsed fraction as large as $\sim 25\%$ and a much richer behavior of the pulse profiles, which can be asymmetric and skewed, as indeed observed in some sources, e.g. the M7  RX J0420.0-5022 \citep{2004A&A...424..635H}.
We found, however, that in these cases the crust tends to become isothermal rather quickly, so that the late time observed properties are controlled by the structure of the magnetic field only. Albeit being based on a simplified envelope model, our thermal maps are in qualitative agreement with those produced by the detailed model of \citet{2020MNRAS.497.2883K}. We finally note that, when studying magnetic field configurations that do not allow long-term heat retention, the study of the coupling of the thermal evolution to the magnetic one may be not necessary. More M7-like isolated NSs are expected to be revealed by eROSITA \citep{2018IAUS..337..112P}, likely including younger, hotter objects. This will allow to further constrain magneto-thermal evolutionary models.

We remark that these configurations are by no means exhaustive, and other options to increase the $\PF$ and to produce non-sinusoidal pulse profiles are possible. Recently, \citet{2020arXiv200103335G} and \citet{2020NatAs.tmp..215I} investigated models in which the field has a strong toroidal component and/or develops instabilities, which are inherently asymmetric, during the evolution. Moreover, the $\PF$ can be enhanced when considering inherently anisotropic emission such as that of a magnetized atmosphere or a condensed surface, \citep[see e.g.][for a review]{2014PhyU...57..735P}, or even more general models with a prescribed beaming pattern \citep[as in e.g.][for isolated NSs and magnetars]{2013MNRAS.434.2362P,2015MNRAS.452.3357G}. In such cases a much higher $\PF$ can be obtained even for quite symmetric temperature distributions. Given that the main goal of this paper is to investigate how magnetothermal evolution impacts on the surface temperature map of a INS, we preferred to limit ourselves to a simple blackbody emission. More detailed emission mechanisms will be considered in future works \citep[as it has been done in e.g.][for \eighteen\ and magnetars, respectively]{2017MNRAS.464.4390P,2020MNRAS.492.5057T} in order to get a  more comprehensive comparison with observed spectra and lightcurves. In this respect, we note that a direct application to observations requires to include interstellar absorption  and the detector response function, although these are likely not to alter dramatically the expected spectral shape and pulse profiles \citep{1996ApJ...473.1067P}, and possible effects due to the NS magnetosphere. Even more important, it would require some information about the source geometry. 

Viewing angles (i.e. the inclination of the line-of-sight and the star magnetic axis wrt the spin axis) are typically hard to determine in radio-quiet sources like the M7, due to the inherent degeneracy which affects spectroscopy alone. A natural way to solve this problem is by exploiting the additional observables provided by polarization measurements, which are already accessible in the optical band and will be available in the near future at X-ray energies with new generation satellites like {\it XPP} \citep{2019BAAS...51g.181J}. A first constraint for the viewing geometry of \eighteen\ has been put from observations in the optical band by comparing the $\PF$ profile with the predicted and the observed polarization fraction as functions of the viewing angles \citep[][]{2017MNRAS.465..492M}. These estimates, however, are based on a purely dipolar field and cannot be directly compared to the model presented in this work. A model of polarized emission taking into account a consistently evolved multipolar field may in the future provide new insights on this matter.

\added{Another issue when dealing with strongly multipolar fields is the computation of the spin period evolution. For example, taking our axisymmetric model and using the usual spin-down formula \citep[e.g.][]{1983bhwd.book.....S}, with an initial period of $10$--$\SI{50}{\milli\second}$ we get period of $\approx \SI{1}{\second}$, quite shorter than the ones measured in the M7 of the same estimated age, $P\sim3-\SI{10}{\second}$ (\eighteen\ has $P\simeq\SI{7}{\second}$). These values may be reproduced only if the initial period is of the same order of the present one (such long initial periods have been proposed, e.g., by \citet{2012A&A...537L...1H} for SXP~1062). Nevertheless, other factors to account for in this respect may be an underestimate of the initial field (reflecting in a too low $B_\text{dip}(0)$), but also that spin-down losses themselves should be computed self-consistently. A complete model taking into account the full structure of the magnetosphere, akin to the one presented by \citet{2006ApJ...648L..51S} for a dipolar field, should be used to assess the effects of the strong multipolar components on the spin-down.}

\replaced{Finally, we warn that our results have been obtained within the present microphysical input of the PARODY code. In particular, a simple crustal equation of state adapted from \cite{2014PhRvL.112q1101G}, see again {DG+20}, and (iron) envelope composition model was used. A different (and more detailed) framework can alter the thermal transport properties, as well as the timescales of the magnetic evolution.
This has a direct impact when drawing comparisons between our results and observations.}
{
Finally, we stress again that our results have been obtained within the present microphysical input of the PARODY code. In particular, a simple crustal equation of state with no chemical stratification (adapted from \cite{2014PhRvL.112q1101G}, see again DG+20) and an iron only envelope model was used. Moreover, we did not consider the contribution to the thermal conductivity by superfluid neutrons, which can reduce the temperature anisotropy in the crust. In this respect, we note that in all our models the crustal temperature barely exceeds $\SI{e8}{\kelvin}$ in the hottest regions, which are not located in the innermost layers of the crust (see e.g. Fig. \ref{fig:Tevo}); hence, we expect this effect not to be substantial. 
This has to be borne in mind when drawing comparisons between our results and observations, which must we kept on a qualitative ground.}

In any case, we expect the deviations related to the microphysics to be \replaced{only}{mostly} quantitative, rather than qualitative, but a complete exploration of the huge number of possibilities is beyond the scope of this work, \added{in which the focus was on the effects of the magnetic field}.

\acknowledgments
Simulations were run at CloudVeneto, a HPC facility jointly owned by the University of Padova and INFN, and at UCL Grace HPC facility (Grace@UCL). The authors gratefully acknowledge the use of both facilities and the associated support services. RT and RT acknowledge financial support from the Italian MUR through grant PRIN 2017LJ39LM. SP has been supported by the Interdisciplinary Scientific and Educational School of Lomonosov Moscow State University ``Fundamental
and Applied Space Research''.

%

\vspace{5mm}





\bibliography{biblio.bib}{}
\bibliographystyle{aasjournal}



\end{document}